\documentclass[fleqn,usenatbib]{mnras}

\usepackage{newtxtext,newtxmath}

\usepackage[T1]{fontenc}

\DeclareRobustCommand{\VAN}[3]{#2}
\let\VANthebibliography\thebibliography
\def\thebibliography{\DeclareRobustCommand{\VAN}[3]{##3}\VANthebibliography}

\usepackage{graphicx}	
\usepackage{amsmath}	

\usepackage{orcidlink}

\usepackage{hyperref}
\hypersetup{
    colorlinks=true,
    linkcolor=blue, 
    urlcolor=cyan 
    }

\title[Disc asymmetry in JWST galaxies]{Disc asymmetry characterisation in JWST-observed galaxies at $1 < z < 4$}

\author[A. Ganapathy et al.]{
Ananya Ganapathy,$^{1}$\thanks{E-mail: a.ganapathy@sms.ed.ac.uk}\orcidlink{0009-0006-1252-206X},
Michael S. Petersen,$^{1}$\orcidlink{0000-0003-1517-3935},
Rashid Yaaqib,$^{1,2}$\orcidlink{0009-0003-9063-1382},
and Carrie Filion$^{3}$\orcidlink{0000-0001-5522-5029}\\
$^{1}$Institute for Astronomy, University of Edinburgh, Royal Observatory, Blackford Hill, Edinburgh EH9 3HJ, UK\\
$^{2}$Department of Physics, United Arab Emirates University, Al Ain, Abu Dhabi, UAE\\
$^{3}$Center for Computational Astrophysics, Flatiron Institute, New York, NY 10010\\
}

\date{Accepted XXX. Received YYY; in original form ZZZ}

\pubyear{2024}

\begin{document}
\label{firstpage}
\pagerange{\pageref{firstpage}--\pageref{lastpage}}
\maketitle

\begin{abstract}
We present a novel technique using Fourier series and Laguerre polynomials to represent morphological features of disc galaxies. To demonstrate the utility of this technique, we study the evolution of disc asymmetry in a sample of disc galaxies drawn from the Extended Groth Strip and imaged by the JWST Cosmic Evolution Early Release Science Survey as well as archival HST observations. We measure disc asymmetry as the amplitude of the of the $m = 1$ Fourier harmonic for galaxies within redshift ranges of $1 < z < 4$ relative to the amplitude of $m = 0$ harmonic. We show that when viewed in shorter rest frame wavelengths, disc galaxies have a higher asymmetry as the flux is dominated by star forming regions. We find generally low asymmetry at rest frame infrared wavelengths, where our metric tracks asymmetry in morphological features such as bars and spiral arms. We show that higher mass galaxies have lower asymmetry and vice versa. Higher asymmetry in lower mass galaxies comes from lower mass galaxies (typically) having higher star formation rates. We measure the relation between disc galaxy asymmetry and redshift and find no conclusive relationship between them. We demonstrate the utility of the Fourier-Laguerre technique for recovering physically informative asymmetry measurements as compared to rotational asymmetry measurements. We also release the software pipeline and quantitative analysis for each galaxy. 
\end{abstract}




\section{Introduction}\label{sec: Intro}

Since the first `spiral nebulae' galaxies were discovered in the 17th century, much research has been dedicated to understanding their morphology and evolution. Morphological classification systems like the Hubble Sequence \citep{Hubble.1926} have stood the test of time as a straightforward proxy to describe galaxy evolution \citep{Buta_etal(2011),Buta_etal(2015)}. Our understanding of such classifications systems have greatly improved due to the rapid advancement of technology. The deep sky surveys by Hubble Space Telescope (HST) were revolutionary as they provided an opportunity to peer into realms never observed before \citep{Buta_etal(2011)}. However, high redshift galaxies in these surveys often cannot be resolved to the level required for detailed classification and appear to have an irregular or peculiar shape leading to a `classification crisis'\citep{vandenBergh(1996),Conselice_(2003)}. Furthermore, owing to the filters available, HST can only investigate galactic structures in the optical wavelength out until $z\simeq 2.8$ beyond which only the UV rest frame wavelength can be probed \citep{Ferreira_etal(2022),Jacob_etal(2023)}. These shorter wavelengths are dominated by young forming stars \citep{Bruzual_Charlot(2003),Jacob_etal(2023)}, which can bias morphological classification.

With the commissioning of the James Webb Space Telescope (JWST) a new era in astronomy dawned. Since becoming operational, JWST has provided an opportunity to reconsider our understanding of galaxy evolution and the physical processes that govern it by leveraging its higher angular resolution and longer wavelength imaging capabilities \citep{Ferreira_etal(2022)}. The advent of novel technology like JWST has also led to an increase in the amount of data at high redshift allowing astronomers to better classify galaxies in the early Universe. While traditional methods like visual classification by individual science teams \citep[e.g.][]{Ferreira.JWST.2023, Kartaltepe_etal(2023), Jacob_etal(2023)} still prevail, newer projects like Galaxy Zoo have emerged that rely on the public to classify images from datasets like the Sloan Digital Sky Survey \citep[SDSS;][]{Lintott_etal2011,Smith_etal(2024)}. To further improve efficiency, artificial intelligence and machine learning models are also being trained on the classifications made through different citizen science projects \citep[e.g.][]{Gordon.etal(2024)}. 

This paper, inspired by the work characterising detailed structure in simulations \citep{Weinberg.etal.2020,Johnson.etal.2023}, presents a novel classification system using a combination of Fourier and Laguerre polynomials. The combination is an effective modelling framework as it compresses large pixel datasets to a few physically meaningful coefficients. The coefficients are a description of a galactic surface brightness distribution. While visual classification becomes more ambiguous when peering into higher redshifts due to decreasing resolution, the Fourier-Laguerre method performs efficiently and reproducibly as it relies directly on quantifiable pixel information. This method is similar to unsupervised machine learning in that it does not require prior training data to provide representative summary information for a given galaxy \citep{Johnson.etal.2023}.

Several studies have already looked into using Fourier series to describe azimuthally dependent morphological features such as bars, spiral arms and asymmetry \citep[e.g.][]{Reichard(2009),Ghosh.etal.2024}. Our model goes a step further by including Laguerre polynomials to represent the radial exponential surface mass density profile of disc galaxies. Early-Universe disc galaxies are crucial to characterise as they provide the key to understanding the dynamical state of the early Universe. While at high redshift the hierarchical construction of galaxies could destroy the fragile discs, observations suggests that many discs exist in the early Universe \citep{Robertson_etal(2023)}. To improve our understanding of disc structures, the focus of this paper is the evolution of disc asymmetry. 

In the past few decades, observational studies suggest that disc asymmetry (measured using $m = 1$ Fourier mode) show a positive correlation with starburst occurrences \citep[e.g.][]{Zaritsky.Rix.1997}. Using a similar method, \citet{Reichard(2009)} also found a strong relation between asymmetry in disc galaxies and the youth of the stellar population present in SDSS galaxies. Following this, \cite{Zaritsky.etal.2013} measured the $m  = 1$ related lopsidedness in a local volume-complete sample from the Spitzer Survey of Stellar Structure in Galaxies (S$^4$G) survey, and concluded that the presence of structure in the dark matter halo could be sufficient to induce observed asymmetries. More recently, \citet{Yesuf.2021} found that disc asymmetries correlated with higher specific star formation rate at a given stellar mass. In this work, we build on these earlier analyses and introduce a novel Fourier-Laguerre expansion framework for image analysis across multiple bands.

The paper is structured as follows. Section~\ref{sec: Methodology} provides context for our model and explains the working of our pipeline. In Section~\ref{sec: Dataset} we detail the datasets and databases referred to when identifying disc galaxies to expand through the pipeline. Section~\ref{sec: Results} presents the main results of the paper, focusing on characterizing the asymmetry of disc galaxies. Section~\ref{sec: Discussion} includes related discussions. The paper concludes with a synopsis of our main results and future work. The Appendices elaborate on several technical points: Appendix~\ref{app: A} introduces the framework of the \textbf{\textit{FLEX}} pipeline; Appendix~\ref{app: B} details the mathematical background of Fourier-Laguerre polynomials and related uncertainty handling; Appendix~\ref{app: C} presents our algorithms for centring and scale length determination integrated into \textbf{\textit{FLEX}}; Appendix~\ref{app: D} addresses the possibilities for where sample selection could affect our results; Appendix~\ref{app: E} is a comparison of \textbf{\textit{FLEX}}-based asymmetry with classical rotational asymmetry; Appendix~\ref{app: F} details analysis of mock galaxies expanded through \textbf{\textit{FLEX}}.

\section{Methodology} \label{sec: Methodology}

We introduce a new pipeline for Fourier-Laguerre EXpansions: \textbf{\textit{FLEX}}. Compared to traditional morphological classifications, \textbf{\textit{FLEX}} retains more galaxy morphology information. We use \textbf{\textit{FLEX}} to measure the asymmetry of galaxies by calculating the Fourier-Laguerre coefficients that best represent a given galaxy. In this section we briefly describe the technique. Further details are located in Appendices: we explain the user experience in Appendix~\ref{app: A}. The  mathematical details and related uncertainties of the Fourier-Laguerre expansions are presented in Appendix~\ref{app: B}. Our treatment of centring and scale length optimisation is described in Appendix~\ref{app: C}. 

Fourier series are the natural basis expansion for describing azimuthal symmetry present in galactic structure. The azimuthal structure of discs exhibits low order multiplicity (denoted $m$), such that Fourier expansions converge quickly (i.e. with fewer than ten harmonics). Previous studies ranging from early automated classification systems \citep[e.g.][]{Odewahn.2002} to more recent efforts \citep[e.g.][]{Ghosh.etal.2024} have mainly focused on a Fourier approach in successive radial annuli to describe galaxy structure and characterise morphological features in disc galaxies such as bars and spirals. In \textbf{\textit{FLEX}}, the user can choose an $m_{\rm max}$ for the Fourier expansion order, setting the maximum multiplicity to be analysed. We choose $m_{\rm max}=1$ in this paper, as explained below, and defer additional multiplicity investigations to future works (cf. Section~\ref{subsec: Future Work}).

In the radial dimension, disc galaxies largely follow an exponential surface brightness profile \citep{Freeman(1970)}, with many also featuring a super-exponential compact bulge in parametric decompositions \citep[e.g.][]{Laurikainen_etal(2016)}. Our expansion technique uses Laguerre polynomials as the radial basis functions. The exponential weighting function of Laguerre polynomials ($G_n(R_{xy})$) closely approximates the expected exponential profile of a typical disc galaxy, making Laguerre polynomials the ideal basis functions for expanding disc galaxies (refer to Appendix~\ref{app: B} for more detail). In \textbf{\textit{FLEX}}, the user chooses a maximum Laguerre polynomial order, $n_{\rm max}$. Higher values of $n_{\rm max}$ result in finer structure resolution; however, the coefficients become increasingly uncertain owing to the spatial resolution of images. We return to this point when discussing uncertainties in Appendix~\ref{app: B}.

By combining Fourier and Laguerre functions to make a two-dimensional basis, we can efficiently model the surface brightness distribution of discs in fewer terms than a corresponding Fourier-annulus expansion, with only physically motivated assumptions. The combination of polynomials enables compression of large pixel datasets to a few {\it physically meaningful} Fourier-Laguerre coefficients that can be used to study features like asymmetries, bars, spiral arms, and rings.These coefficients can be mathematically described as

\begin{equation}
    \begin{split}
        \hat{c}_{mn} &= \frac{1}{2\pi}\sum_x \sum_y \Sigma(R_{xy}, \phi_{xy})G_n(R_{xy}) \cos\left(m\phi_{xy}\right)\\
        \hat{s}_{mn} &= \frac{1}{2\pi}\sum_x \sum_y \Sigma(R_{xy}, \phi_{xy})G_n(R_{xy}) \sin\left(m\phi_{xy}\right)
    \end{split}
\label{eq:6}
\end{equation}
where $\Sigma(R_{xy}, \phi_{xy})$ is the surface brightness of a given pixel. Further description, including the definitions of the polynomials $G_n$, is in Appendix~\ref{app: B}.

We restrict our analysis in this work to asymmetries, which are the $m=1$ set of coefficients. Furthermore, these coefficients are also unaffected by the inclination of the galaxy. As shown by several past studies \citep[e.g.][]{Reichard(2009),Amvrosiadis_etal.(2024)} $m = 1$ has widely been used to track the asymmetry of a galaxy's surface density. We further distil the radial information in the coefficients into a single metric measuring the strength of asymmetry in a galaxy:
\begin{equation}
    A_m = \sqrt{\sum_{n_{\min}}^{n_{\max}} \left( c_{mn}^2 + s_{mn}^2 \right)}\Bigg/\sqrt{\sum_{n_{\min}}^{n_{\max}} \left( c_{0n}^2\right)}.
    \label{eq:7}
\end{equation}
All $s_{0n}$ terms in the numerator are zero. Physically $A_{m}$ quantifies the Fourier deviation from the exponential profile, in units emphasising the significance of the deviation (in gravitational potential energy) from an axisymmetric disc. The normalisation term in the denominator is analogous to the the total luminosity of the galaxy measured in native image units, here being pixels. Refer to Appendix~\ref{app: B} for detailed mathematical explanation.

\begin{figure}
	\includegraphics[width=\columnwidth]{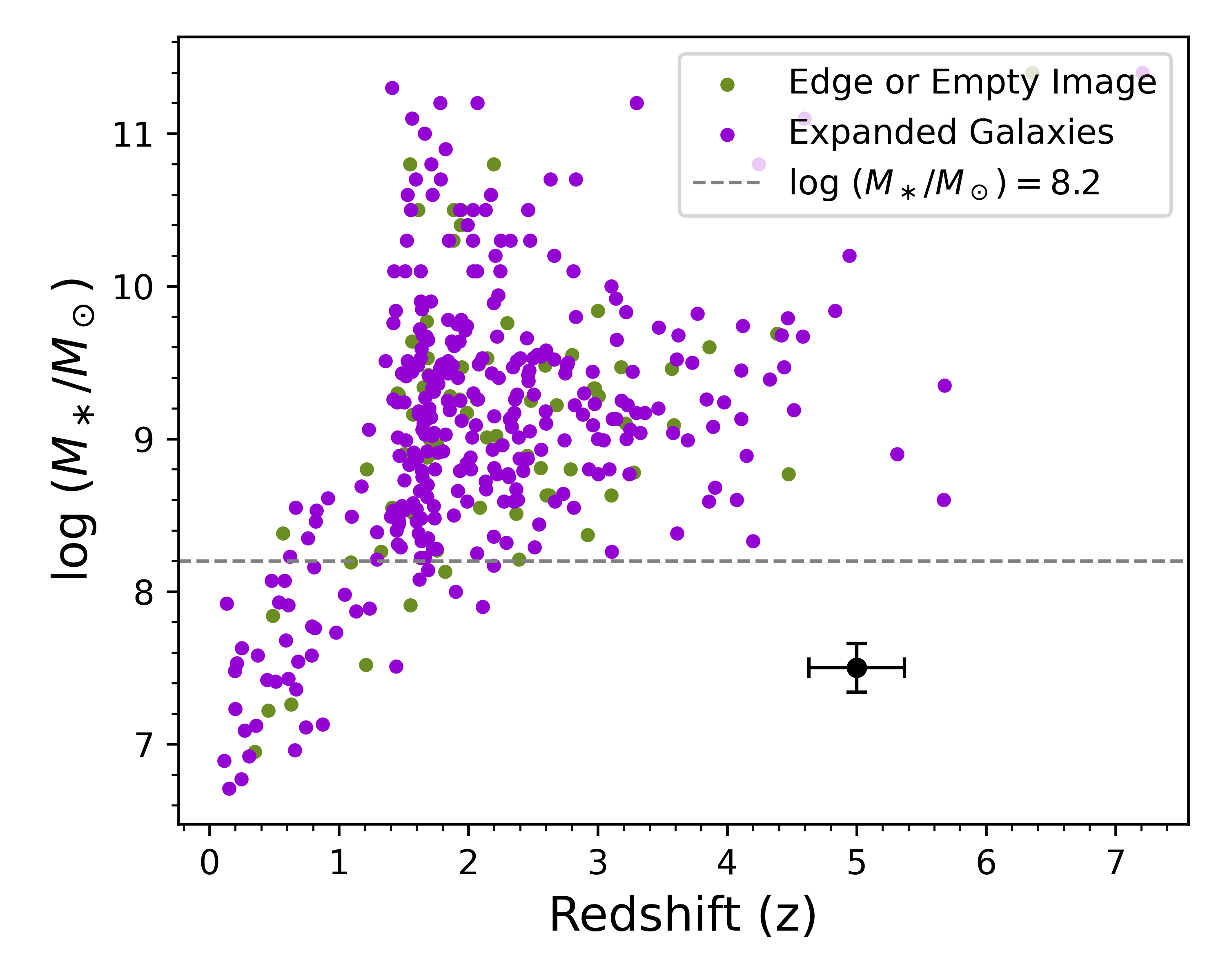}
    \caption{Distribution of 387 classified disc galaxies redshift values against their respective stellar mass as reported in \textit{CANDELS} \citep{Stefanon_2017}. The green points correspond to those that have been discarded either due to empty image arrays or being situated at the edge of the FITS image. The remaining violet data points correspond to the 271 successfully expanded galaxies. All galaxies lying over the threshold of $M = 10^{8.2}{\rm M}_\odot$ are treated as the primary sample (Refer to Section~\ref{sec: Results}). The black symbol in the lower right corner indicates the median uncertainty in redshift and stellar mass determinations for the primary sample.}
    \label{fig:dataset}
\end{figure}
 
\section{Data}\label{sec: Dataset}

For this paper, we expand disc galaxies that lie in the Extended Groth Strip (\textit{EGS}). These galaxies were imaged by the JWST Cosmic Evolution Early Release Science Survey \citep[\textit{CEERS};][]{Finkelstein_2022} Near-Infrared Camera (NIRCam) Fields 1 - 3 and 6. We use the data reductions from the \textit{CEERS} team \citep{Finkelstein_2022} as well as the re-reductions of the Cosmic Assembly Near-IR Deep Extragalactic Legacy Survey (\textit{CANDELS}) HST imaging \citep{Grogin_2011, Koekemoer_2011} of the same fields. Each galaxy was expanded in  F444W, F410M, F356W, F277W, F200W, F115W NIRCam JWST  filters. These were then complemented with HST Wide Field Camera 3 (WFC3) filters F125W and F160W and HST Advanced Camera for Survey (ACS) F606W and F814W. This gives a total of ten filters per galaxy in our sample. In the re-reduction from the \textit{CEERS} team, all images are aligned and matched in pixel scale. \textit{CEERS} imaging these \textit{EGS} galaxies is complemented heavily by the extensive auxiliary data in \textit{CANDELS}.

We use previously (visually) classified disc galaxies from the \citet{Ferreira.JWST.2023} database in this work\footnote{The database is hosted on GitHub and can be found \href{https://github.com/astroferreira/CEERS_EPOCHS_MORPHO/tree/main}{here}. Nonparametric galaxy properties are computed using the `Morfometryka' code \citep{Ferrari_2015}, which we use in Appendix~\ref{app: E}.}. These were selected by identifying galaxy IDs labelled `disc' class and with a `True' confidence label. In this database, any galaxy bearing a `True' confidence meant that the majority class is unanimous amongst the six classifiers \citep{Ferreira.JWST.2023}. From the 3956 galaxies analysed by \citet{Ferreira.JWST.2023}, 531 were classified as discs. Out of these 531, 144 galaxies were from the UltraDeep Survey (\textit{UDS}) and the remaining 387 come from the \textit{EGS}. To ensure the homogeneity of our sample, particularly in derived quantities, we focus on the 387 \textit{EGS} galaxies. We discuss possible sample selection biases in our data sample in Appendix~\ref{app: D}. Figure~\ref{fig:dataset} summarises the 387 classified \textit{EGS} disc galaxies by plotting stellar mass\footnote{Obtained from the compiled \textit{EGS} catalogue of stellar masses, \texttt{M\_med}, and related uncertainties, \texttt{s\_med} \citep{Stefanon_2017}.} vs. best redshift\footnote{Obtained from compiled \textit{EGS} catalogue of redshifts {\tt zbest} \citep{Stefanon_2017}. All redshift measurements are photometric estimates; the median uncertainty is $\sigma_z=0.37$. This is small enough so as not to appreciably change any of our conclusions. The redshift estimates are consistent with more recent estimates \citep{Kodra.2023}.}. The region between redshift 1 and 2 is densely populated while most galaxies appear to have a stellar mass ranging from $ 10^{7}{\rm M}_\odot < {\rm M}_{\ast} <10^{10} {\rm M}_\odot$. To reduce sample bias, we treat galaxies with ${\rm M}_{\ast} >10^{8.2}{\rm M}_\odot$ as our main sample for interpreting trends and discuss results for lower mass galaxies throughout. 

For application in this paper, \textbf{\textit{FLEX}} is designed to expand galaxies within a 110 by 110 pixel ($3.3 \times 3.3 $ arcseconds) postage stamp cutout from the \textit{CEERS}-reduced FITS Image in each filter, by adopting the centre pixel coordinates provided in the \textit{CANDELS} catalogue. However, \textbf{\textit{FLEX}} has been developed to adapt to any pixel cutout size as the expansion coefficients are not dependent on the window size. To improve the accuracy of the expansion, all non-central galaxies within this cutout are masked\footnote{For 24 galaxies there was foreground contamination due to multiple galaxies lying close to the main galaxy. These have been removed for the purpose of this paper, and they will be further explored in future work (cf. Section~\ref{subsec: Future Work}).}. This automated process begins by estimating a scale radius for the galaxy from the centre of the cutout. This is done by fitting an exponential profile to the F444W image using \texttt{scipy.optimize.curve\_fit} and the \textit{CANDELS}-provided estimate for the centre. Once the initial exponential scale radius has been determined in F444W, we define a mask radius that is six times the exponential scale radius and keep it constant for all other filters. Outside this radius, pixels that have flux above three times the median absolute deviation of the pixel intensity within the postage stamp are turned to \texttt{NaN} values. This technique, called sigma clipping, is thus used to ensure no contamination from foreground or background sources. Within the defined radius of the galaxy of interest, \textbf{\textit{FLEX}} expands all the pixels, but only expands those outside the radius that fall below the threshold.

We summarises the uncertainties related to this masking process in Appendix~\ref{app: B}. After applying the mask, we then proceed to optimise our determination of the centre of the galaxy and tune the scale length, as detailed in Appendix~\ref{app: C}. In the end, our final sample consists of 271 successfully expanded galaxies as shown by the violet points in Figure~\ref{fig:dataset}\footnote{During data validation, 81 galaxies were discarded due to either having empty image arrays or lying within 50 pixels of the edge of the detector, reducing the available background for the expansions. These excluded galaxies are represented by the green points in Figure~\ref{fig:dataset}. Besides this, 11 galaxies were eliminated as at the reported \textit{CANDELS} coordinates, a disc galaxy was not visible.}. In the following section, the primary sample of 271 galaxies reduces due to sigma clipping and mass cuts applied. In the following section, for each reduced sample we mention the total number of galaxies in the analysis.

\section{Results} \label{sec: Results}

\begin{figure}
	\includegraphics[width=\columnwidth]{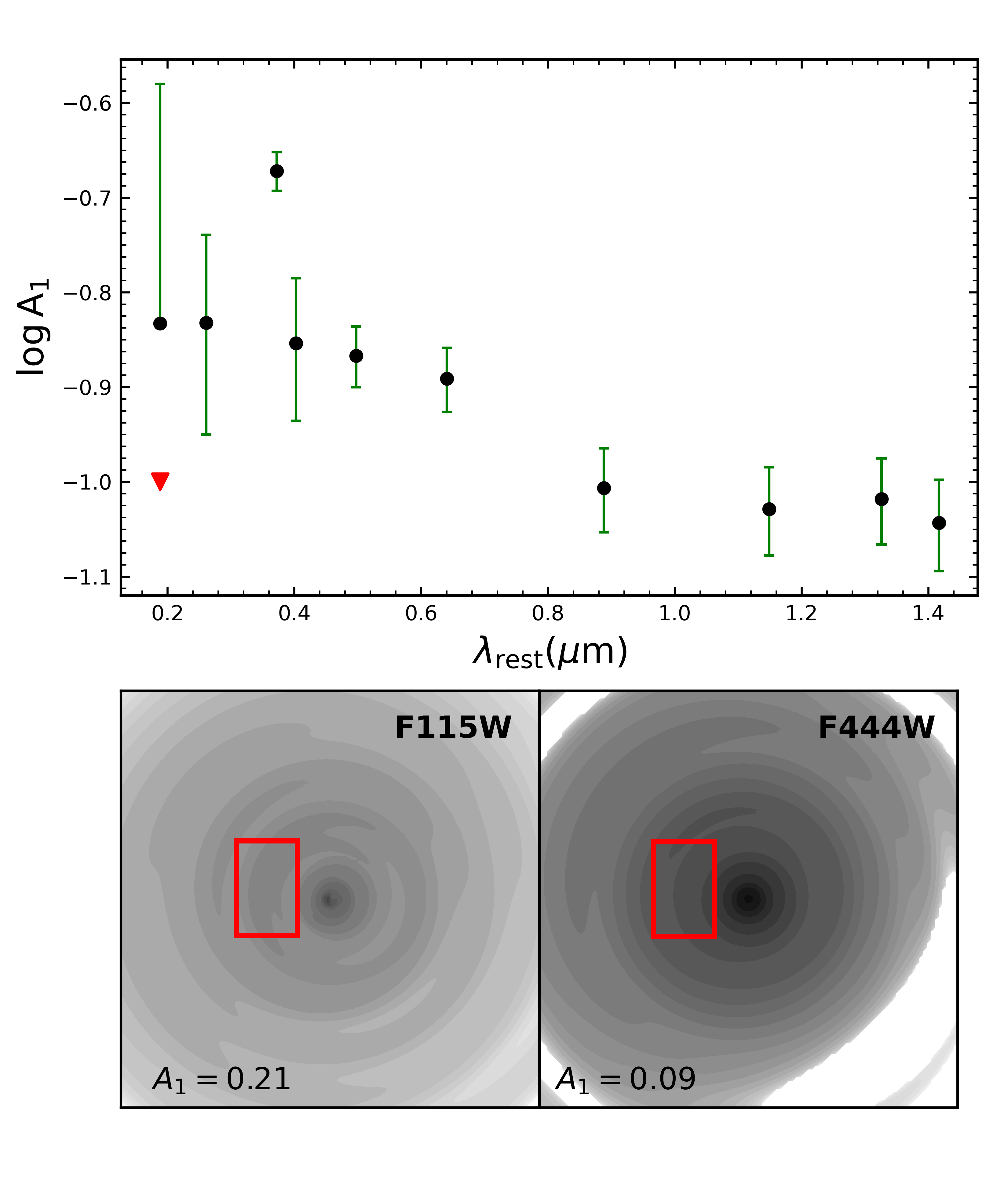}
    \caption{Top panel: Log Asymmetry of typical galaxy, EGS23205 ($\log(M_\ast/M_\odot) = 11.2 $, SFR = $64.83 M_\odot / \text{yr}$), as a function of rest frame wavelength (assuming $z=2.07$). The inverted red triangle denotes that the lower limit on the lower error bar for the asymmetry measurement made in F606W extends to negative infinity (zero asymmetry). Bottom panels: (Left) Expanded surface density ($m = 1$) of EGS23205 in F115W with clumps of new stellar formation visible. (Right) Expanded surface density ($m = 1$) of EGS23205 in F444W filter, where the galaxy has a lower asymmetry value. The highlighted red box focuses on a clump of star formation present in the spiral arms visible in the F115W filter (left) but appearing smoother in the F444W filter (right). Compare with higher-order expansions in Figure~\ref{fig:expansion_grid}, which further resolve the star-forming region outlined by the red box.}
    \label{fig:main_plot_1}
\end{figure}

Given our \textbf{\textit{FLEX}}-derived asymmetry metric, we investigate the asymmetry of galaxies (1) as a function of observed wavelength, (2) as a function of redshift, and (3) as a function of galaxy mass and specific star formation rate\footnote{We use \textit{EGS} Catalogue of galaxy physical properties {\tt 6a\_tau} star formation rate \citep{Stefanon_2017}, which is derived using an exponentially declining star formation model. Using other estimated star formation rates does not change our conclusions.}.

\subsection{A pedagogical example} \label{subsection:pedagogical_example}

As a pedagogical demonstration of the asymmetry calculation pipeline, Figure~\ref{fig:main_plot_1} focuses at how (the log of) asymmetry depends on rest frame wavelength for EGS23205 ($z = 2.07^{+0.17}_{-0.02}$, $\log(M_\ast/M_\odot)~=~11.2 $,~SFR~=~$64.83 M_\odot / \text{yr}$) expanded across 10 JWST and HST filters (Refer to Section~\ref{sec: Dataset} for filter names). This galaxy is one of the highest redshift barred disc galaxies discovered to-date \citep{Guo.CEERS.2023}. 

The asymmetry in EGS23205 is higher (relative to the median asymmetry value of this galaxy) at shorter wavelengths, where emission from young stars is expected to dominate \footnote{However, short of 0.4$\mu$m, the uncertainties increase owing to low overall flux from the galaxy, cf. Figure~\ref{fig:expansion_grid} and discussion in Appendix~\ref{app: F}.}. The phenomenon of increasing asymmetry at shorter wavelengths is commonly observed in the multiwavelength expansions performed here (see next Section). For EGS23205, the clump-like structures present in the spiral arms at shorter wavelength (the bottom left of Figure~\ref{fig:main_plot_1}) are not visible in the longer wavelength (the bottom right of Figure~\ref{fig:main_plot_1}). This smoother appearance in the longer wavelengths results in a lower computed asymmetry value relative to the median value. The highlighted regions in the bottom panels of Figure~\ref{fig:main_plot_1} focus on a star forming region present in a spiral arm. Visual inspection of these regions also suggest that the clumps of star formation present in the F115W filter (left) appear smoother in the F444W filter (right).

The $A_{1}$ value in JWST F115W ($\lambda_{\text{rest}} = 0.37 \mu m$) is high (relative to the median value) owing to prominent star formation clumps resolved in the image. The same clumps are also observed in the HST F606W ($\lambda_{\text{rest}} = 0.19 \mu m$) and F814W ($\lambda_{\text{rest}} = 0.26 \mu m$) filters, but the low signal-to-noise in HST images creates significant uncertainty in the measurement.

Figure~\ref{fig:expansion_grid} shows how EGS23205 appears in the complete suite of JWST and HST filters (upper row), with the corresponding \textbf{\textit{FLEX}} expansion in the lower row. Unlike the bottom panel of Figure~\ref{fig:main_plot_1} which is the expanded surface density for the $m = 1$ mode, Figure~\ref{fig:expansion_grid} explores the expansion with $m_{\rm max} = 8$ mode and radial order $n_{\rm max}~=~24$. Hence, higher $m_{\rm max}$ is useful when trying to understand morphological features like the clumpy appearance of star formation as highlighted by the red box in the bottom panel of Figure~\ref{fig:expansion_grid}. This region focuses on the same star forming clump highlighted in Figure~\ref{fig:main_plot_1}. The galaxy is well-reconstructed where the detection is significant (i.e. near the centre of the image), but aliasing appears in the outskirts. This aliasing is purely visual and does not affect our conclusions.

\begin{figure*}
    \includegraphics[width=\linewidth]{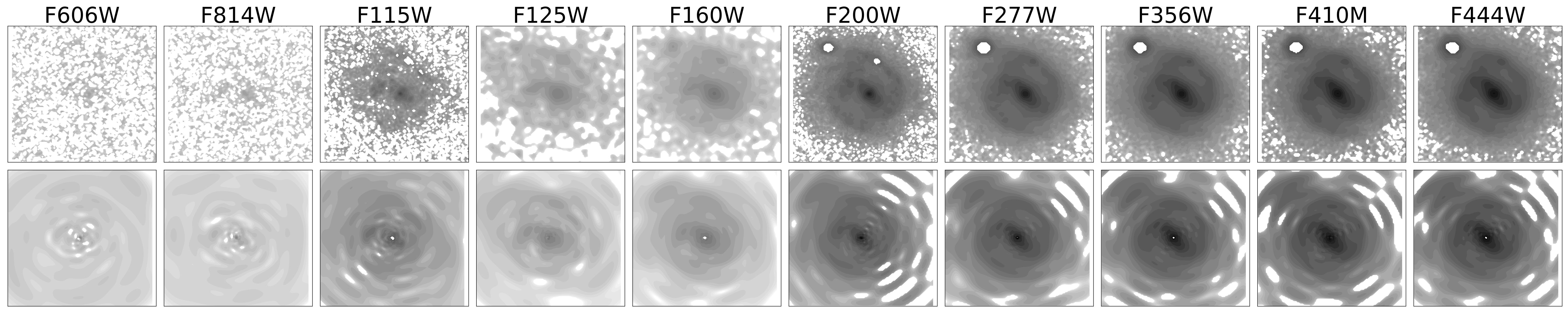}
    \caption{Surface Density of EGS23205 (Top Row) and the reconstructed surface density (Bottom Row) in JWST filters (F444W, F410M, F356W, F277W, F200W, F115W) and HST filters (F160W, F125W, F814W, F606W). These expansions are conducted with $m_{\rm max} = 8$ mode and radial order $n_{\rm max}~=~24$. The top row also demonstrates the background masking procedure (Refer Section~\ref{sec: Dataset} for detailed explanation) applied within a 110 by 110 pixel bound. The red boxes in the bottom panel represent the same clump of star formation from Figure~\ref{fig:main_plot_1}.}
    \label{fig:expansion_grid}
\end{figure*}

\begin{figure}
	\includegraphics[width=\columnwidth]{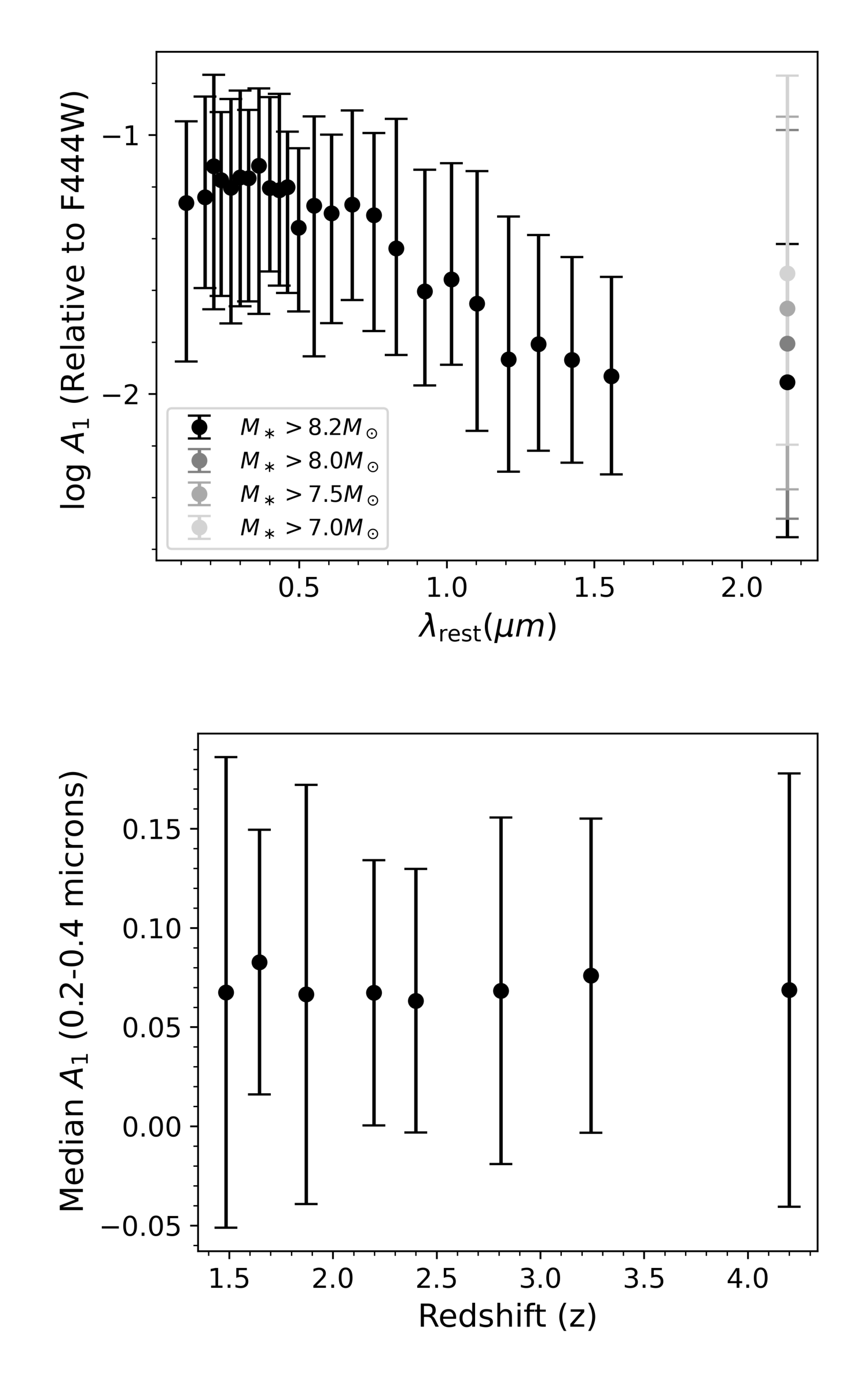}
    \caption{(Top) Log asymmetry of 243 galaxies equally distributed in 25 rest frame wavelength bins for galaxies with stellar mass $> 10^{8.2} {\rm M}_\odot$. The inverse relation suggests that asymmetry appears to be higher in bluer filters where stellar formation is visible. (Bottom) Median asymmetry of 243 galaxies (stellar mass $> 10^{8.2} {\rm M}_\odot$) in 8 equally distributed redshift bins. Measured in the UV rest frame region ($0.2 < \lambda_{\text{rest}} < 0.4\mu m$), there appears to be no relation between redshift and asymmetry. In both panels, the error bars denote the 1$\sigma$ range of the observations, not the uncertainty on the mean.}
    \label{fig:main_plot_2}
\end{figure}

\subsection{Asymmetry vs. wavelength and redshift}

The top panel of Figure~\ref{fig:main_plot_2} compares the (log) median asymmetry in 25 equally distributed rest frame wavelength bins for 243 galaxies expanded through \textbf{\textit{FLEX}}. We assume that asymmetry in longer wavelengths (those measured in F444W filter here) is due to `structural asymmetry'\footnote{That is, asymmetry that stems from asymmetry in the mass, rather than the light distribution; refer to Section~\ref{subsec: Structural Asymmetry} for further discussion.}, and $A_1$ in this filter has been subtracted from the asymmetry measurement of other filters. This allows us to only investigate the evolution of `star formation asymmetry'\footnote{That is, asymmetry owing to the light distribution, not necessarily reflected in the mass distribution. See Section~\ref{subsec:starformationasymmetry} for further discussion.} in disc galaxies. Once the bins have been determined and equally populated, the median $A_{1}$ value and error bars are determined in each bin. Here, the error bars represent the $1 \sigma$ range of values and not the uncertainty on the mean. To better visualise the underlying trends, we show the log of all median $A_{1}$ values, with uncertainties computed in linear space and plotted as the log values. The strong inverse relationship (R$^{2}=0.887$) suggests that the asymmetry of galaxies appears higher at shorter wavelengths. This is analogous to Figure~\ref{fig:main_plot_1} and Figure~\ref{fig:expansion_grid} which confirms that at shorter wavelengths the determined asymmetry metric is probing new clumpy stellar formation. The longer wavelengths (last bin) in Figure~\ref{fig:main_plot_2} is dominated by low stellar mass galaxies located at low redshifts as seen in Figure~\ref{fig:dataset}. Hence, Figure~\ref{fig:main_plot_2} quantifies the change in median asymmetry for the last bin when considering different mass cuts. Given the practicalities of the mass sample we have, galaxies with a stellar mass $< 10^{8.2} {\rm M}_\odot$ were excluded (grey points in Figure~\ref{fig:main_plot_2}).

The bottom panel of Figure~\ref{fig:main_plot_2} investigates the evolution of median asymmetry in 8 equally distributed redshift bins. This has been recorded in the UV rest frame region ($0.2 < \lambda_{\text{rest}} < 0.4\mu m$) as star formation is expected to dominate the emission. A similar mass cut ($ > 10^{8.2} {\rm M}_\odot$) has also been applied to the sample in the bottom panel of Figure~\ref{fig:main_plot_2}. The lack of observed trend (R$^{2}=0.006$) suggests that the asymmetry of disc galaxies does not evolve with increasing redshift in the \textit{CEERS} sample, although larger mass-limited samples are required to confirm this trend. 

\begin{figure}
	\includegraphics[width=\columnwidth]{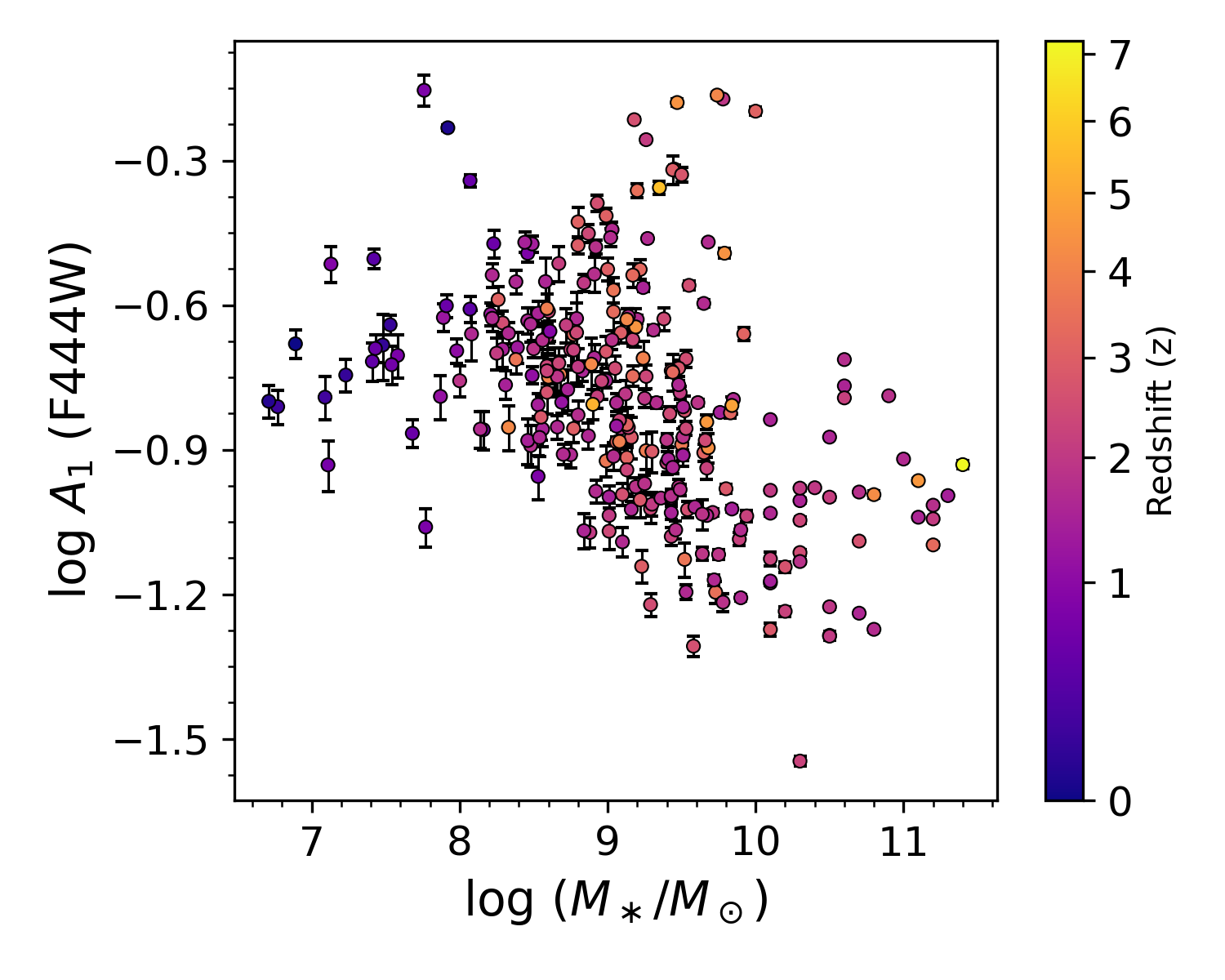}
    \caption{Stellar Mass as a function of log$A_{1}$ for 271 galaxies expanded through \textbf{\textit{FLEX}} in the JWST F444W filter. Each galaxy is coloured with respect to its redshift values as reported in the \textit{CANDELS} catalogue \citep{Stefanon_2017}. This plot investigates the inverse relationship between structural asymmetry and the stellar mass of a galaxy.}
    \label{fig:main_plot_3}
\end{figure}

\subsection{Asymmetry vs stellar mass and star formation}

Figure~\ref{fig:main_plot_3} shows galaxy stellar mass against log$A_{1}$, uniformly measured in the F444W filter. Here, the asymmetry metric is probing structural asymmetry measured in the longer rest frame wavelengths. For our set of \textit{CEERS} disc galaxies, we find that lower mass galaxies have higher associated structural asymmetry. 

The structural asymmetry uncertainty is higher for smaller galaxies (refer to Appendix~\ref{app: B} for further explanation). Each galaxy in Figure~\ref{fig:main_plot_3} has also been coloured based on its redshift value as reported in \textit{CANDELS} \citep{Stefanon_2017}. Given this for a fixed $A_{1}$ value, higher redshift galaxies tend to have higher stellar mass as shown in Figure \ref{fig:dataset}. However, studies also suggest that this could be a result of selection bias as at higher redshifts, low stellar mass galaxies become more difficult to detect \citep{Castro-Rodriguez_etal(2012)}. 

\begin{figure}
	\includegraphics[width=\columnwidth]{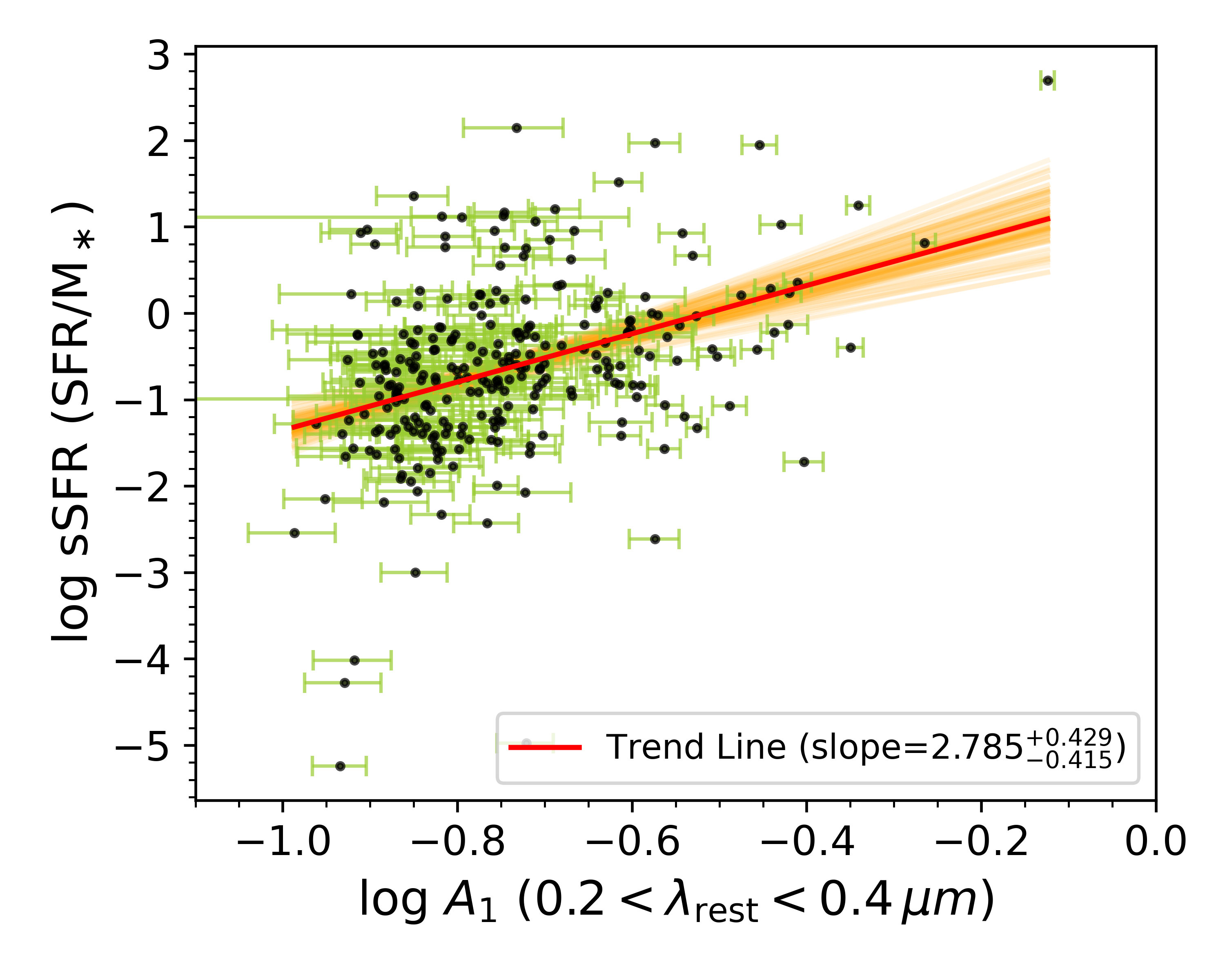}
    \caption{Disc Asymmetry of all 259 galaxies against their specific star formation rate \citep[sSFR][]{Stefanon_2017} in log-space as measured between $0.2 <~\lambda_{\text{rest}} < 0.4\mu m$. This figure indicates that our defined asymmetry metric tracks stellar formation in the UV rest frame region.}
    \label{fig:main_plot_4}
\end{figure}

Figure~\ref{fig:main_plot_4} shows the relationship between specific star formation rate (sSFR) and asymmetry as measured by our metric in the UV rest frame region ($0.2~<~\lambda_{\text{rest}}~<~0.4\mu m$) for 259 galaxies. Here, sSFR is defined as the ratio of SFR to stellar mass. Unlike Figure~\ref{fig:main_plot_3}, the asymmetry metric here is directly investigating asymmetry in young stars as this portion of the spectrum is dominated by star formation. To reduce the effect of outliers when recovering the trend, asymmetry and sSFR values outside the 3 sigma region were clipped and hence the total sample of 271 galaxies reduced to 259. The linear positive trend in Figure~\ref{fig:main_plot_4} was fit to the trimmed data using \textit{emcee}. The initial guesses for slope and y intercept was $[0.5, 0.5]$ and were restricted within the bounds of -10 to 10. To find the optimal parameters, 32 walkers were introduced and each simultaneously explored the parameter space for 5000 iterations. The calculated slope is $2.79_{-0.415}^{+0.429}$. Assuming that emission in the rest frame UV corresponds primarily to star formation, Figure~\ref{fig:main_plot_4} asserts that higher asymmetry corresponds to star formation as determined with respect to sSFR. 

\section{Discussion} \label{sec: Discussion}

\subsection{Asymmetry as a Dynamical Classification Tool}

\subsubsection{Mergers driving structural asymmetry}\label{subsec: Structural Asymmetry}

Mergers are widely known to affect galaxy asymmetry and are a consequence of the hierarchical formation of structure predicted by $\Lambda$-cold dark matter model \citep[e.g.][]{Bottrell.2024}. 

The impact of mergers on SFR has widely been researched and observations show that mergers can trigger star formation \citep{Pearson_etal(2019)}, which would appear in the shorter wavelength regime. In addition to possibly forming new stars, mergers can distort the shapes of galaxies and build their stellar masses. Stars in galaxies can be traced back to either of two origins: {\it in situ} (stars formed from gas within the galaxy it belongs to) or {\it ex situ} (stars formed outside of the host galaxy in a system that has since merged, \citealt{VRodriguez-Gomez_etal(2016)}). Hydrodynamic cosmological simulations such as the Illustris Project \citep{Vogelsberger(2014), Genel_etal(2014)} have been extensively used to study the role of {\it in situ} and {\it ex situ} stars in galaxy evolution. Studies using this suite of simulations \citep{Lackner_etal.(2012),VRodriguez-Gomez_etal(2016),Pillepich_etal(2018)} suggest that {\it ex situ} star formation prevails largely in galaxies with stellar mass $ > 10^{11} {\rm M}_\odot$. Galaxies with stellar mass $ < 10^{11} {\rm M}_\odot$ are still affected by {\it ex situ} star formation, however the overall fraction of {\it ex situ} stars is small \citep[see][their Figure 6]{VRodriguez-Gomez_etal(2016)}. These observations are also dependent on redshift with lower redshift galaxies having higher {\it ex situ} star formation fractions. However, these results could be biased as with increasing redshift it is harder to resolve the effect of merger activity on disc galaxies. Referring back to Figure \ref{fig:dataset}, it appears that only 7 galaxies lie above this threshold ($ > 10^{11} M_{\odot}$) with most galaxies heavily populated within the range of $ 1 < z < 3$. In the \textit{CEERS} disc-selected galaxies, {\it in situ} star formation appears dominant thus indicating that our metric is not tracking major merger activity. This conclusion is supported by examining the rotational asymmetry values from \citep{Ferreira.JWST.2023}, which indicate that none of the galaxies in our sample would have been classified as mergers in the Concentration, Asymmetry, and Smoothness (CAS) system (see next subsection)\footnote{Our sample purposely selects against strong merger-driven asymmetry, as we use the sample of confidently classified discs from \citet{Ferreira.JWST.2023}. See discussion in Appendix~\ref{app: D}.}.

\cite{Bottrell.2024} measured rotational asymmetry for a sample of simulated galaxies in TNG, finding that minor (mass ratio between 0.1 and 0.25) and `mini' (mass ratio between 0.01 and 0.1) mergers can drive structural asymmetry that lives on up to 3 billion years after merging. Those authors studied this by determining the offset in the Star Forming Main Sequence as it relates to the symmetry of galaxies. We call the longest wavelength measurement of the asymmetry the `structural asymmetry' measurement, as even up to $z=3$, the F444W filter probes the infrared wavelength region, where the stellar light has a strong component from lower-mass stars that trace the long-lived stellar population. The \textbf{\textit{FLEX}}-derived asymmetry measure, at first interpretation, does suggest that some galaxies may have significant asymmetry in these longer wavelengths. These galaxies are part of the lower mass sample that was excluded from the primary analysis. For all galaxies in the primary analysis, the mean asymmetry is $\langle A_1\rangle_{\rm F444W}=0.15$.

\subsubsection{Relationship to the CAS system}
\label{subsubsec:casrelationship}

`Asymmetry' has, for more than two decades, meant the 180$^\circ$ rotational asymmetry metric from the CAS system \citep{Conselice_(2003)}. To compute the rotational asymmetry, a square pixel cutout from an image is rotated by 180$^\circ$, and then the asymmetry metric is computed as the total absolute difference between the original and rotated image, normalised by the original image. In general, rotational asymmetry from the CAS system will emphasise distortions in lower surface brightness features (i.e. tending towards the outskirts) where the absolute deviations can be the largest, which is beneficial for identifying broad galaxy classifications. The utility of the rotational asymmetry measurement as a classifier was explored in detail in \citet{Conselice.2000}, where the metric was calibrated to a nearby sample. This work followed earlier applications to some of the then-highest redshift observations from HST \citep{Schade.1995,Abraham.1996}, who established algorithmic measurements as being as reliable as visual classifications at high redshift. Rotational asymmetry has often been used as a merger indicator \citep{Conselice_(2003)}, particularly in tandem with measures of the concentration of stellar light. When combined with galaxy colour, rotational asymmetry can distinguish spirals, ellipticals, and mergers \citep{Conselice.2000}.

Our definition of an asymmetry metric is distinct from rotational asymmetry, but fully contains the rotational asymmetry information. We retain many of the best qualities of rotational asymmetry measurements (e.g. low dependence on resolution) owing to the global nature of the basis functions. Additionally, we have a key advantage in our algorithmic centre-finding scheme that incorporates the technology used to define the metric, giving robust uncertainty estimates.

In Appendix~\ref{app: E} we compare our asymmetry measure directly to the CAS system. We do not recover any obvious relationship between the two metrics; in particular at asymmetry $A_1\ge0.3$ measured either via rotation or \textbf{\textit{FLEX}}, we find no similarities between galaxies, suggesting that the two systems are in fact probing different properties. Tests on mock images (Appendix~\ref{app: F}) suggest that the differences owe to the effective spatial scales of asymmetry probed by the two metrics: rotational asymmetry will emphasise the asymmetries that encompass the largest pixel area, while \textbf{\textit{FLEX}}-derived asymmetries will capture the strongest asymmetries relative to an unperturbed galaxy\footnote{\textbf{\textit{FLEX}} can, in principle, be made to reproduce the rotational asymmetry metric by choosing a different weighting scheme for the radial coefficients that determine the asymmetry value.}. \textbf{\textit{FLEX}} has the added benefit of connecting to dynamics through simulations performed in a similar framework \citep[e.g.][]{Petersen.exp.2022,Petersen.linearresponse.2024}, enhancing the interpretability of the \textbf{\textit{FLEX}}-derived asymmetries. We discuss the utility of \textbf{\textit{FLEX}} measurements for a standard use of rotational asymmetry -- detecting mergers -- in the next Section.

\subsection{Asymmetry Resulting from Star Formation}
\label{subsec:starformationasymmetry}

Stellar evolution models show that emission at wavelengths within $0.2 < \lambda_{\text{rest}} < 0.4\mu$m is dominated by star formation \citep[e.g.][]{Bruzual_Charlot(2003)}. In this region of Figure~\ref{fig:main_plot_1} and \ref{fig:main_plot_2}, the calculated asymmetry is relatively higher as compared to longer wavelengths. As highlighted in Section \ref{sec: Intro}, asymmetry is known to track star formation which can be visually confirmed by the bottom left panel of Figure~\ref{fig:main_plot_1}. Hence the observed asymptomatic behaviour in Figure~\ref{fig:main_plot_1} and ~\ref{fig:main_plot_2} can be associated with the youth of stellar populations found within the galaxies. In  particular, these clumps of stellar formation are present in the spiral arms as visually identified in Figure~\ref{fig:main_plot_1} and Figure~\ref{fig:main_plot_2}.

Figure~\ref{fig:main_plot_4} shows the relationship between sSFR and asymmetry as measured by our metric in  the UV rest frame region ($0.2<~\lambda_{\text{rest}}~<~0.4\mu m$). Assuming that emission in the rest frame UV corresponds primarily to star formation, higher asymmetry in disc galaxies in the UV likely traces star formation, and is thus correlated with higher sSFR. The Spearman correlation coefficient for this relation is 0.378 (p-value = $3.639 \times 10^{-10}$) indicating a weak positive relation between asymmetry and sSFR.  

Redder filters, which probe the longer rest frame wavelengths, are advantageous when viewing larger morphological features like bars \citep[i.e. an $m = 2$ Fourier feature;][]{Ghosh.etal.2024} and spiral arm structure. This is because these filter sets trace the old redder stellar population, enabling the study of smooth broader galactic features. Despite not tracking young, blue stellar populations in the redder filters (e.g. the JWST F444W filter), our determined metric confirms that there exists underlying structural asymmetry in disc galaxies (Figures~\ref{fig:main_plot_1} and~\ref{fig:main_plot_2}). In addition, when viewed in the near IR region, asymmetry and sSFR appear to have a negative correlation. Thus it can be assumed that our metric is recognising another asymmetry contributor that impacts the redder stars without inducing additional star formation (Refer to Section \ref{subsec: Other Sources of Asymmetry} for more information).

\subsubsection{Star formation and low stellar mass: a highly asymmetric regime}
\label{subsubsec:lowstellarmassasymmetry}

Figure~\ref{fig:main_plot_3} shows an inverse relation between asymmetry and stellar mass suggesting that low stellar mass galaxies have higher asymmetry. This result makes sense in context with other works looking at asymmetry and star formation rate relationships. Recent research indicates that for galaxies in the \textit{CEERS} dataset, a direct relation between a galaxy's stellar mass and sSFR can be drawn \citep{Cole_etal(2023)}. A similar study using the JWST Advanced Deep Extragalactic Survey (JADES) also concluded that low mass galaxies experience more recent active star formation between z$\sim 4 - 9$ \citep{Simmonds_etal(2024)}. At lower redshift, observations utilising the Sydney–AAO Multi-object Integral field spectrograph (\textit{SAMI}) Galaxy Survey show that low mass galaxies have proportionately more  H$\alpha$ gas \citep{Bloom(2017)}. This is a strong indicator of star formation as it traces the ionised gas of star forming H II regions \citep{Tacchella_etal(2022)}. Besides this, earlier analysis utilising SDSS demonstrated that low mass galaxies are more prone to experience star bursts and young stellar formation \citep{Kauffmann_etal(2003)}. These observations can be used to support the inverse relationship seen in Figure~\ref{fig:main_plot_3}. Additionally, \citet{Reichard(2009)} deduced that lower-mass galaxies tend to exhibit greater disc asymmetry (measured by the $m = 1$ mode) and appeared more lopsided. Building on this, \citet{Yesuf.2021} introduced a data-driven framework to identify galaxy properties correlated with specific star formation rates. Asymmetry, as computed in the CAS system by \citet{Bottrell.2019}, was found to correlate with the position of a galaxy relative to the sSFR main sequence as a function of galaxy stellar mass.

\subsection{Other Sources of Asymmetry} \label{subsec: Other Sources of Asymmetry}

The asymmetry metric, defined in this study, is physically informative as it describes the surface brightness profile from the galactic centre to the outskirts (see comparison to the CAS system, which emphasises asymmetry in the outskirts, in Section~\ref{subsubsec:casrelationship}). The physical causes of asymmetry in lower-mass galaxies seems to be explained by the relationship with star formation (Section~\ref{subsubsec:lowstellarmassasymmetry}), and our asymmetry metric is not as sensitive to mergers as the CAS system, which means that other physical processes may contribute to asymmetry in our sample. Hence, beyond merger-driven asymmetry and star formation-driven asymmetry, we consider other causes of asymmetry that have been theorised: (1) Active Galactic Nuclei (AGN) activity, (2) dark matter halo (sub)structure, and (3) cosmological accretion. 

 AGN have been proposed as drivers of asymmetry, usually owing to their relationship with star formation. Previous observational studies have shown that radio-mode AGN feedback quenches star formation as the energy released causes shock waves and heat which prevent clouds of gas and dust from cooling down \citep{ACFabian(2012)}. For galaxies in SDSS, \citet{Reichard(2009)} found a direct relation between asymmetry and measured power of centrally hosted AGN, suggesting that highly asymmetric galaxies may tend to host relatively more powerful AGN and hence quench star formation more. However, past research focusing on AGN within \textit{CANDELS} galaxies have found that AGN do not affect galaxy asymmetry when compared to control galaxies \citep{Villforth_etal(2014)}\footnote{A similar study focusing around z $\sim$ 2 found minor asymmetries  in 44$\%$ of AGN-hosting galaxies \citep{Kocevski_etal(2012)}.}.

Long-lived dark matter halo structure has been shown in simulations to create disequilibrium and asymmetries in the stellar disc, which can persist for Gyr \citep{Grand.2023,Johnson.etal.2023}. Some of the structure may be induced by mergers or interactions, while other halo structure owes to natural evolutionary modes that are present in many halos \citep{Weinberg.2023}. Such asymmetries should be present at all wavelengths, as they will affect both the older stellar population as well as where star formation may be enhanced. Work connecting observed disc asymmetries to halo structure necessitates further theoretical development, but if conclusively demonstrated, disc asymmetry could provide a powerful tool to probe the dynamical state of dark matter halos as well as their substructure.

Finally, the accretion of gas along filaments can produce asymmetry. N-body simulations investigating this asymmetry suggest that cold gas accretion can form a thin, kinematically cold, lopsided disk that lives on as the disc galaxy evolves \citep{Bournaud_etal.(2005)}. In this scenario, the asymmetry will show up in largely younger stars, but would also show up in any gas distribution. Future observations of either gas structure or kinematics can help determine whether gas accretion can sustain observed asymmetries in discs.

\subsection{Future Prospects for \textbf{\textit{FLEX}}}\label{subsec: Future Work}

As discussed in Section~\ref{sec: Methodology}, \textbf{\textit{FLEX}} has currently been set up to measure disc asymmetry using the $m = 1$ mode related coefficients. However, we aim to focus future work on using higher order Fourier modes ($m = 2, 3...$) to characterise morphological features such as bars and spirals after accounting for the effect of inclination and galaxy ellipticity. 

Besides this, \textbf{\textit{FLEX}} only expands a single galaxy within a cutout while masking background galaxies. We aim to extend its functionality to performing multiple galaxy expansions within a given pixel bound. This improvisation can be used to study inter-galactic dynamics such as mergers using the coefficients produced. 

After testing our methodology with mock galaxies (see Appendix~\ref{app: F}), we find that \textbf{\textit{FLEX}} produces coefficients with no direct dependence on the galaxy image resolution (and therefore redshift-dependent resolution effects). Future research could leverage this feature to morphologically classify high-redshift galaxies, providing insights into the formation of the earliest disc galaxies.

Additionally, \textbf{\textit{FLEX}} also lends itself to studying galaxies in the Local Group by leveraging higher-order harmonics. This is advantageous, as higher-order harmonics can help describe finer features of galaxies, such as accretion around black holes.  

\section{Conclusion}  \label{sec: Conclusion}

In this work, we characterise the asymmetry of 271 disc galaxies lying within the \textit{EGS} field imaged by JWST in \textit{CEERS} and archival HST observations. For this work, we expand previously identified disc galaxies \citep{Ferreira.JWST.2023} within redshift ranges of $1~<~z~<~4$ using the \textbf{\textit{FLEX}} pipeline. This novel framework uses Fourier-Laguerre polynomials to compress pixel information into coefficients that describe galaxy morphology. Compared to traditional nonparametric classification methods, \textbf{\textit{FLEX}} retains information on both the exponential profile of a disc galaxy as well as physically informative deviations. In this work, we use \textbf{\textit{FLEX}} to measure the asymmetry of a galaxy using the set of $m=1$ radial coefficients. The $m=1$ radial coefficients describe the asymmetry (lopsidedness) of galaxy light distributions. The main conclusions of our work are as follows:

\begin{enumerate}
    \item Asymmetry is higher at shorter rest frame wavelengths (upper panel of Figure~\ref{fig:main_plot_2}). This is because in bluer filters, clumps of young stellar population dominate the emission (bottom left panel of Figure~\ref{fig:main_plot_1}). Redder filters can be used to probe asymmetry of features like bars and can be used to determine the underlying structural asymmetry of disc galaxies (bottom right panel of Figure~\ref{fig:main_plot_1}). 
    
    \item Asymmetry does not evolve with redshift in the UV rest frame region (bottom panel of Figure~\ref{fig:main_plot_2}). However, a larger mass limited sample is needed to find a conclusive relation between asymmetry and redshift. 
  
    \item Disc galaxies with lower stellar mass are more asymmetric and vice versa in the F444W filter (Figure~\ref{fig:main_plot_3}). This suggests that the structural asymmetry, determined at longer wavelengths, is inversely proportional to the stellar mass of a disc galaxy. This aligns with earlier analyses showing that lower stellar mass galaxies have higher presence of star formation rate \citep{Cole_etal(2023), Simmonds_etal(2024)}. We also conclude that for a given $A_{1}$ value, higher stellar mass galaxies appear to be located at high redshifts. 
    
    \item The asymmetry of a disc galaxy is positively correlated with sSFR in the UV rest frame region (Figure~\ref{fig:main_plot_4}). We interpret this as direct evidence that $A_1$ is indicative of stellar formation happening within the disc galaxy. 
\end{enumerate}

\section*{Acknowledgements}

We thank the anonymous referee for their feedback that helped us improve this paper. AG acknowledges financial support from the University of Edinburgh Physics and Astronomy Career Development Scholarship. MSP acknowledges support from a UKRI Stephen Hawking Fellowship. MSP and CF thank the EXP collaboration for feedback on an early version of this work. This work is based in part on observations made with the NASA/ESA/CSA James Webb Space Telescope and NASA/ESA Hubble Space Telescope obtained from the Space Telescope Science Institute, which is operated by the Association of Universities for Research in Astronomy, Inc., under NASA contract NAS 5–26555. These observations are associated with program(s) CANDELS. The authors of this paper utilised the following software: numpy \citep{harris2020array}, scipy \citep{2020SciPy-NMeth}, HDF5 \citep{F.Brand(1998)}, Matplotlib \citep{Hunter:2007}, Astropy \citep{astropy:2013, astropy:2018, astropy:2022}, emcee \citep{Daniel_Foreman-Mackey}, and lintsampler \citep{Naik.lintsampler.2024}. 

\section*{Data Availability}

FITS Images for this project were taken from the \textit{CEERS} dataset available \href{https://ceers.github.io/index.html}{here}. \textit{CANDELS} data, used in this paper, is also publicly available \href{https://archive.stsci.edu/hlsp/candels/egs-catalogs}{here}. \textbf{\textit{FLEX}}, created by the authors, can be accessed \href{https://github.com/ananyag8303/FLEX-pipeline}{here}.

\bibliographystyle{mnras}
\bibliography{reference} 



\appendix

\section{\textbf{\textit{FLEX}} Pipeline framework} \label{app: A}

\textbf{\textit{FLEX}} is an automated pipeline which can accept a list of user input Extended Groth Strip (\textit{EGS}) galaxy IDs and returns postage stamps, expansion coefficients, and asymmetry measurements. The process begins with locating the desired galaxy within the provided FITS images using reported (RA,~Dec) values from the \textit{CANDELS} catalogue \citep{Stefanon_2017}. If a galaxy cannot be located, it is reported to the user and saved to a `unclassifiable' log file. Once located, a 110 by 110 pixel postage stamp of the galaxy is created centred around the given (RA,~Dec). To reduce interference, bright pixels outside the desired galaxy's radius are masked using sigma clipping. Refer to Section~\ref{sec: Dataset} for detailed explanation about masking. A first estimate of the scale radius of the galaxy is determined using \texttt{scipy.optimize.curve\_fit} to fit an exponential profile for the surface brightness profile of the galaxy.
Following this estimate, we check that the galaxy image is not at the edge of the FITS file or empty. If it meets either of the conditions, the galaxy details are appended to the same `unclassifiable' log file.

Once all filter image details are saved onto an HDF5 \citep{F.Brand(1998)} formatted file, the centre and scale length for the expansion (Appendix~\ref{app: C}) are determined. These values are then set as parameters in the Fourier-Laguerre coefficients calculations. The determined Fourier-Laguerre coefficients are saved onto the same HDF5 file as the filter images upon completion. To visualise the expanded surface density, the pipeline displays a contour map of the expanded surface density (Equation~\ref{eq:5}) alongside an amplitude chart for the filter the galaxy was expanded in. 

\section{\textbf{\textit{FLEX}} Uncertainty Handling}\label{app: B}

\textbf{\textit{FLEX}} represents galaxies using a Fourier-Laguerre 2D expansion, which has also been used in \citet{Weinberg.etal.2020,Johnson.etal.2023}. We detail the expansion technique and define the $A_1$ asymmetry metric in this Appendix.

Disc galaxies tend to have an exponential surface brightness profile
\begin{equation}
    \Sigma(R) \propto \exp\left(-\frac{R}{a}\right)
    \label{eq:1}
\end{equation}
where $R$ is the 2D radius and $a$ is the scale length of the disc. 

In the radial $R$ coordinate, Laguerre polynomials are used as the exponential weighting function matches the exponential profile of a typical disc galaxy (cf. Equation~\ref{eq:1}):

\begin{equation}
G_n(R)=\frac{1}{a\sqrt{n+1}}~\exp\left(-\frac{R}{a}\right)~L_n^1\left(\frac{2R}{a}\right).
\label{eq:2}
\end{equation}

Here, $L_n^1$ is the associated Laguerre polynomial of order $\alpha=1$ and degree $n$, and $a$ is the expansion scale length (chosen to be equivalent to the disc scale length). As the $n=0$ polynomial is a constant, Equation~\ref{eq:2} matches the expected exponential surface brightness (Equation~\ref{eq:1}). Increasing values of $n$ adds additional `nodes' to the Laguerre function, which correspond to excesses (decrements) above (below) the exponential profile.

Galaxies are often reasonably smooth in the azimuthal coordinate $\phi$, with low-order multiplicity $m$ that can be described by Fourier series. When combined, Fourier-Laguerre basis functions given by
\begin{equation}
    \begin{split}
        f_{mn} &= G_n(R)\exp\left(im\phi\right),
    \end{split}
\end{equation}
can accurately model the surface brightness profile of discs in two dimensions, in appropriately chosen combinations. In practice, we compute the exponential using cosine and sine, corresponding to the real and imaginary part of the expansion, respectively. Determining the weighting of each function in the expansion is straightforward, from the two-dimensional surface brightness of the galaxy (Equation~\ref{eq:6})
where the discrete summations are over individual pixels in the image, indexed by $(x,y)$. The reconstructed surface density at each $(x,y)$ pixel location, $\hat{\Sigma}(R_{xy}, \phi_{xy})$, is then
\begin{equation}
\begin{split}
        \hat{\Sigma}(R_{xy}, \phi_{xy}) = \sum_{m=0}^{m_{\rm max}}&\sum_{n=0}^{n_{\rm max}} G_n(R_{xy})~\times\\
        &\left[\hat{c}_{mn}\cos\left(m\phi_{xy}\right) + \hat{s}_{mn}\sin\left(m\phi_{xy}\right)\right].
        \label{eq:5} 
    \end{split}
\end{equation}
In practice, the expansion is truncated at some maximum Fourier multiplicity $m_{\rm max}$ and a maximum radial order $n_{\rm max}$, set by the quality of the data or by the user. We defer a detailed exploration of the effect of expansion order to a future work (cf. Section~\ref{subsec: Future Work}). In the remainder of the work, we assume coefficients (Equation~\ref{eq:6}) are estimated from pixels, and drop the $\hat{\cdot}$ notation.

\begin{figure}
	\includegraphics[width=\columnwidth]{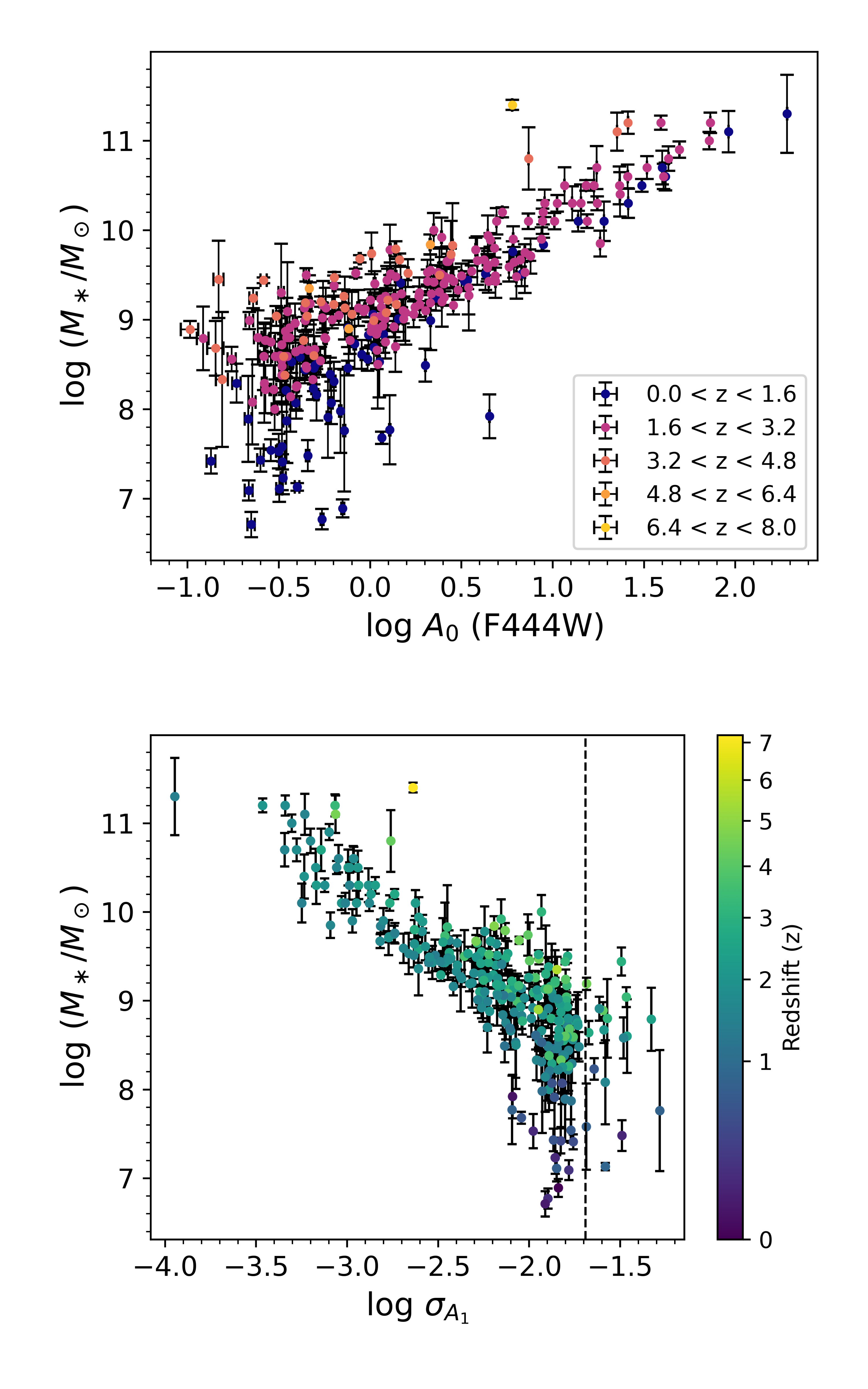}
    \caption{(Top) Scatter Plot displaying a positive linear relationship between log $A_{0}$ and Stellar Mass \citep{Stefanon_2017} as measured in F444W filter. This relationship appears to be analogous to the Luminosity - Mass relationship as $A_{0}$ describes the total luminosity of the galaxy. This is because, $A_{0}$ is the lowest order term and thus carries the highest significance in the expansion. (Bottom) Scatter Plot comparing the evolution of $A_{1}$ metric uncertainty with Stellar Mass as reported in \textit{CANDELS} \citep{Stefanon_2017}. The points are also coloured based on their redshift values obtained from \textit{CANDELS} \citep{Stefanon_2017}. This figure asserts that higher mass galaxies have lesser $A_{1}$ uncertainty associated for a fixed redshift value. The vertical dashed line represents the point beyond which the uncertainty of the expansion is dominated by centre finding and truncation uncertainty.}
    \label{fig:combined_plot}
\end{figure}

To analyse morphological features, the coefficients are quantified by computing the total amplitude $A_{m}$ (for $m>0$) in a harmonic order, normalised by the axisymmetric amplitude as shown in Equation~\ref{eq:7}. The top panel of Figure~\ref{fig:combined_plot} demonstrates that the expected Luminosity-Mass relationship can be retrieved through plotting the normalising axisymmetric amplitude in the JWST F444W filter, i.e. the denominator of Equation~\ref{eq:7} against the derived stellar mass.

In the main text we restrict our analysis to the $A_1$ term, and will study higher harmonics in future work (Refer Section~\ref{subsec: Future Work}). We refer to the $A_1$ term as the `disc asymmetry'. The bottom panel of Figure~\ref{fig:combined_plot} describes the inverse linear relationship between the uncertainty on disc asymmetry and stellar mass. For a fixed stellar mass, at higher redshift it appears that there is higher associated uncertainty in $A_{1}$. When considering high mass galaxies, they appear more luminous and have a prominently defined galactic centre. Due to this, the model can more accurately predict the $A_{0}$ metric. Hence, the other $A_{m}$ metrics such as $A_{1}$ have lower associated uncertainty as shown in the bottom panel of Figure~\ref{fig:combined_plot}.

We treat four sources of uncertainty in the calculation of $A_1$: (1) formal image uncertainty; (2) expansion truncations; (3) centre related uncertainty; (4) masking related uncertainty.

To treat formal image uncertainty (1), we create random noise realisations of the images from the reported pixel uncertainty and pass those images through the pipeline, calculating the median and standard deviation of $A_1$. We also use an empirical estimate for the effect of image uncertainty on the calculation of $A_0$ (used in the calculation of $A_1$) as a function of the background relative to the galaxy peak surface brightness (see Appendix~\ref{app: F}).

To treat expansion truncation uncertainty (2), for each galaxy, we perform expansions with varying maximum Laguerre order $n_{\rm max}$ and record the asymmetry values $A_1$ for each expansion. With increasing $n_{\rm max}$ the value of $A_1$ converges. The $n_{\rm max}$ value after which $A_1$ shows least variance informs the choice of the order that is required for each galaxy. In the convergence regime, the standard deviation of $A_1$ about the mean is recorded as the expansion truncation uncertainty. For all galaxies considered here, beyond the median radial order of 12 the value of $A_1$ converges. 

The uncertainty related to the expansion centre (3) is set by the tolerance level during the minimisation procedure described in Appendix~\ref{app: C}. The uncertainty in this case is the tolerance set of 0.01, giving a floor in the systematic uncertainty for $A_1$.

The uncertainty corresponding to background galaxy masking (4) has been accounted for in two ways: error due to galaxy radius scale factor chosen and error from threshold set for sigma clipping. Tests conducted showed that decreasing the scale factor increases the asymmetry value while the increasing the threshold for sigma clipping increases the asymmetry value. Both these errors contribute to less than 10\% of a difference in values and are not a dominant source of uncertainty in the \textbf{\textit{FLEX}} pipeline. 

Taken together, the image related and centre finding uncertainty are the most dominant sources and are represented by the vertical dashed line in the bottom panel of Figure~\ref{fig:combined_plot}. It can be concluded that lower mass galaxies are more dominated by these systematic errors as compared to higher mass galaxies.

\section{Centring and Scale Length Optimisation}\label{app: C}

In Section~\ref{sec: Dataset}, we use the \textit{CANDELS}-reported centre to create the galaxy postage stamp. To refine our estimates for the centre and maximise the fidelity of the reconstruction, we follow a two-step procedure to (1) minimise the large-scale asymmetry, and (2) select the scale length that minimises the amplitude of the higher-order axisymmetric coefficients relative to the lowest-order axisymmetric coefficient.

To determine the centre, we define an algorithmic centre for the galaxy based on minimising the large-scale asymmetry. In practice, the centre of a galaxy is determined by minimising the a radial-term-limited dipole metric using the Newton - Raphson method,
\begin{equation}
    A_1=\sqrt{\sum_{n_{\rm min}}^{n_{\rm max}} \left(c_{1n}^2 + s_{1n}^2\right)}.
    \label{Raphson}
\end{equation}
Here, $n_{\rm min}$ and $n_{\rm max}$ are the lowest and highest radial orders considered, respectively for the expansion. When determining the centre, a smaller $n_{\rm max}$ ($n_{\rm max} = 10 $) is chosen to minimise the dipole on larger scales, while we maintain $n_{\rm min}=0$, as in the main text. Hence, \textbf{\textit{FLEX}} reiterates over Equation~\ref{Raphson} for each pair of pixel coordinates till it reaches the `central' coordinates that minimise large-scale asymmetry. 

Unlike fitting elliptical isophotes, this dynamic method allows the algorithm to account for non-axisymmetric features in the centre calculation. When compared to the RA and Dec values reported in the \textit{CANDELS} catalogue, we find an offset with respect to the calculated centres.

\begin{figure}
	\includegraphics[width=\columnwidth]{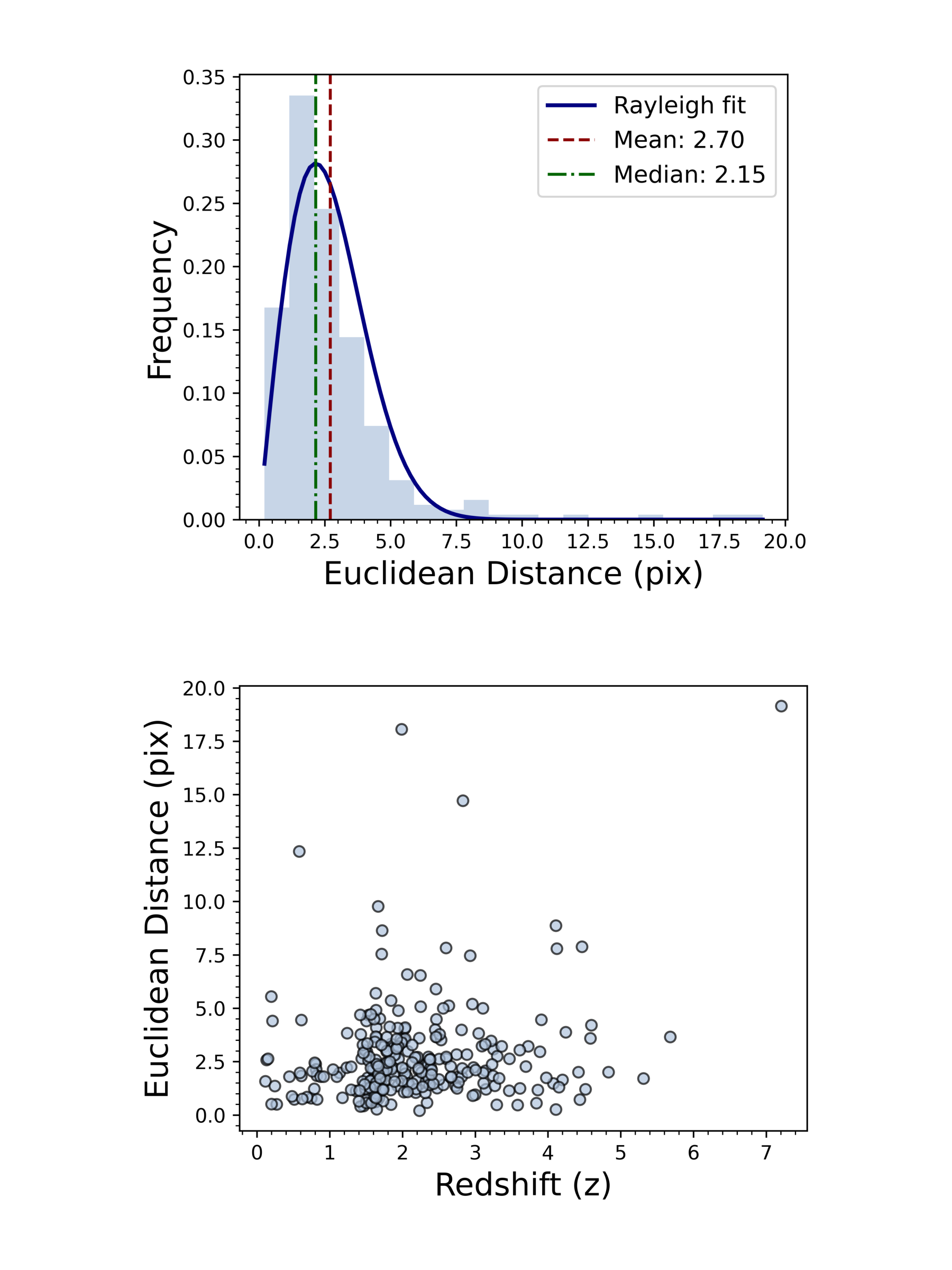}
    \caption{(Top) Rayleigh distribution of the offset in calculated centre values in F444W filter relative to values reported in the \textit{CANDELS} catalogue \citep{Stefanon_2017}. This distribution suggests that the calculated offsets are random variables. (Bottom) Scatter plot comparing the centroid offset in the F444W filter with its corresponding redshift value for each galaxy. The absence of a trend in this distribution suggests that our determined centre is not redshift dependent when compared to \textit{CANDELS}-computed centres.}
    \label{fig:euclidean_dist}
\end{figure}

To quantify the offset and determine the cause, we measure the Euclidean distance between the reported and calculated centres were determined in each filter (F444W, F410M, F356W, F277W, F200W, F115W, F814W, F606W, F125W and F160W). The results for F444W filter are shown in Figure~\ref{fig:euclidean_dist}. Assuming that the calculated centres ($x,y$) are independent variables with standard normal distributions having $\mu = 0$ and $\sigma \in (0,\infty)$, the Euclidean distance is simply the magnitude of vector ($x,y$). The most appropriate distribution to quantify a random `offset' is a Rayleigh distribution as depicted in the top panel of Figure \ref{fig:euclidean_dist}. Meanwhile, the lower panel of Figure \ref{fig:euclidean_dist} is a scatter plot between the Euclidean distance for each galaxy in F444W filter along with its corresponding redshift value. The random scatter of data in the lower panel of Figure~\ref{fig:euclidean_dist} suggests that the offset calculated is unrelated to redshift. This further proves that \textbf{\textit{FLEX}} isn't affected by redshift and can accurately track morphological features throughout the Universe. While Figure \ref{fig:euclidean_dist} only reports the results for the F444W filter, these observations are consistent across all filters.

To optimise the scale length, we minimise
\begin{equation}
    \text{ratio} = \frac{\sqrt{\sum_{n=1}^{n_{\rm max}} c_{0n}^2}}{c_{00}}
    \label{Scale_Length}
\end{equation}
where $c_{00}$ represents the zeroth order axisymmetric (monopole) coefficient for $m,n = 0$. In physical terms, this is akin to the total pixel intensity or luminosity of the galaxy. Similarly, $c_{0n}$ is the axisymmetric coefficient only in the radial direction as $m$ is fixed to 0. An optimal scale length for the expansion is one that minimises Equation~\ref{Scale_Length}. Physically, Equation~\ref{Scale_Length} measures how the luminosity of a galaxy spreads radially and can be used to determine when the light goes from being centrally concentrated to being extended radially.  

\textbf{\textit{FLEX}} is structured such that this is initially found in the F444W filter before keeping it fixed for all the other filters. This is because F444W filter has the longest effective wavelength of $4.35 \mu m$, giving the best determination of the mass distribution of the galaxy. Hence, \textbf{\textit{FLEX}} calculates the Fourier - Laguerre coefficients for each scale length value before identifying the value that minimises Equation~\ref{Scale_Length}.

\section{Sample Selection and possible bias}\label{app: D}

Section~\ref{sec: Dataset} describes the dataset used to expand disc galaxies. As we utilised only unanimously agreed upon disc galaxies in the \citet{Ferreira.JWST.2023} database \footnote{The database is hosted on GitHub and can be found \href{https://github.com/astroferreira/CEERS_EPOCHS_MORPHO/tree/main}{here}.}, it is possible that our dataset is biased towards a particular class of disc morphology. For reference, we only analysed EGS galaxies bearing labels `disc' and `True' from the database created by \citet{Ferreira.JWST.2023}. This was because any classification bearing a `False’ indicator meant that at least one of the six visual classifiers in \citet{Ferreira.JWST.2023} disagreed with the majority classification. While selecting only confident morphological discs ensures consistency in classifications, the number of disc galaxies identified within the \textit{CEERS} Image Releases is likely underestimated, and may be biased towards lower stellar mass galaxies with lower structural asymmetry.  

On the other hand, \citet{Ferreira.JWST.2023} states that disc galaxies were visually identified by looking for features such as concentrated bulges and disc envelopes alongside more prominent features such as spiral arms and bars. As the authors note, these features are harder to recognise at higher redshift and could be confused with spheroids. This potentially biases the classifications at higher redshift where image resolution is reduced, with the inclusion of spheroidal galaxies, which tend to have lower rotational asymmetry \citep{Conselice_(2003)}.

Similarly as the inclination of a galaxy increases, a disc galaxy can appear barred and features such as spiral arms are not visible when edge on. This could lead to ambiguity when trying to classify between galaxy morphology types and lead to a biased dataset. \citet{Bournaud.etal.2005} also notes that when a disc galaxy is not observed face on, fitting Sersic profiles is not a reliable classification method. This is because the surface density profile is no longer exponential and the range over which the light is well-fit by a Sersic profile is reduced as the bulge covers the disc. Hence, the deviation of a galaxy profile from an exponential profile can bias our initial step of background masking and coefficient determination. Furthermore, when observing disc galaxies edge on, its radial luminosity profile could be altered due to dust attenuation.

\section{Comparison with rotational asymmetry}\label{app: E}

\begin{figure}
	\includegraphics[width=\columnwidth]{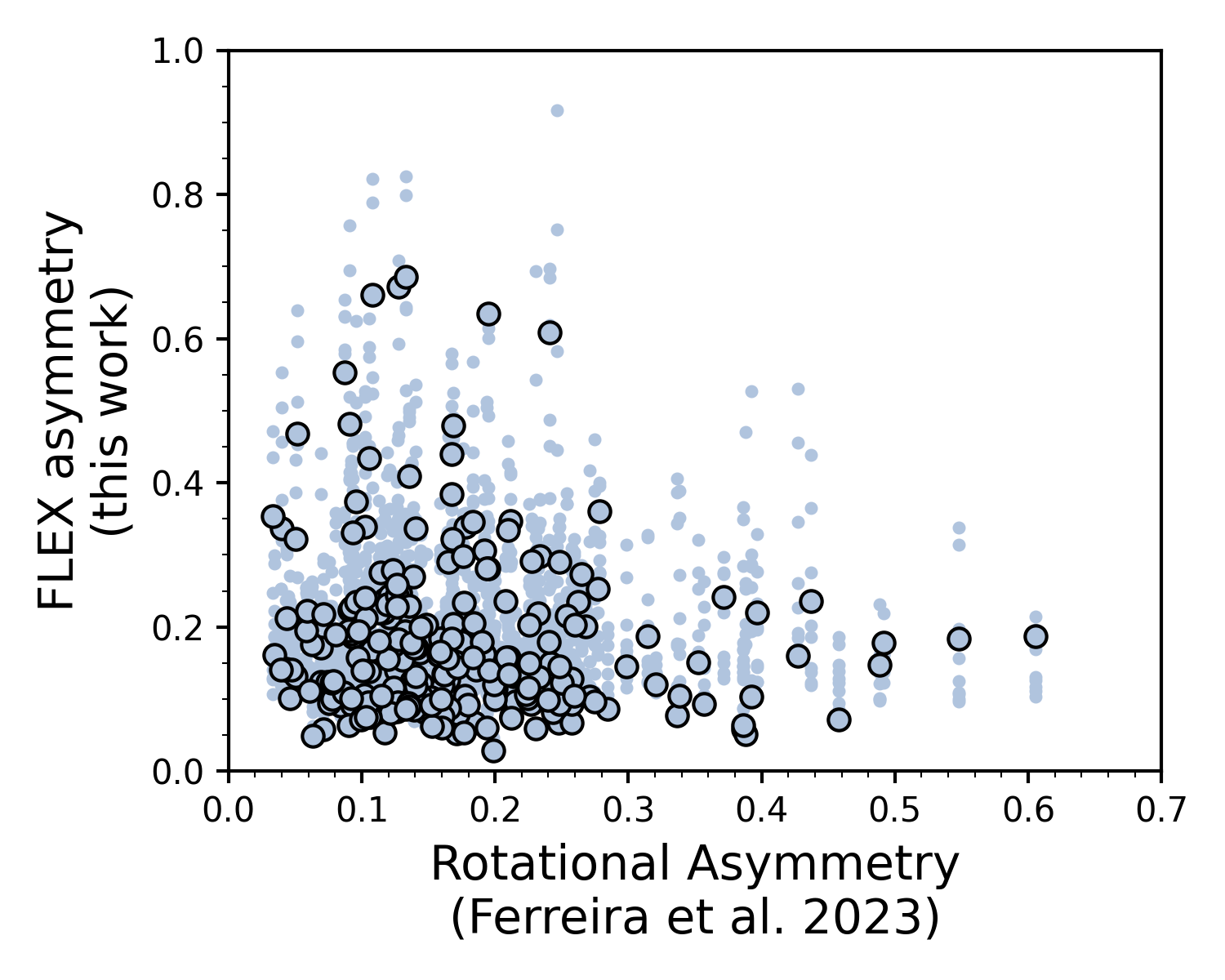}
    \caption{Comparison of rotational asymmetry \citep{Ferreira.JWST.2023} and \textbf{\textit{FLEX}}-derived asymmetry. \citet{Ferreira.JWST.2023} measurements are made in restframe optical bands per galaxy; for the \textbf{\textit{FLEX}}-derived asymmetry we show the value in each filter, demonstrating that in no filter do we find mapping between the metrics. }
    \label{fig:rotational_asymmetry}
\end{figure}

In Section~\ref{subsubsec:casrelationship}, we qualitatively discuss the relationship between our defined asymmetry metric (Equation~\ref{eq:7}) and the rotational asymmetry metric discussed in detail in \citet{Conselice_(2003)}. 

Figure~\ref{fig:rotational_asymmetry} shows a direct comparison of rotation asymmetry from \citet{Ferreira.JWST.2023} and our \textbf{\textit{FLEX}}-derived asymmetry. In no regime do we find obvious agreement, and we find particular disagreement at high asymmetry values in either metric. Our \textbf{\textit{FLEX}}-based procedure is more similar to residual asymmetry \citep[e.g.][]{Bottrell.2024}, which is supported by our conceptually similar results (e.g. the relationship of asymmetry with star formation).

The difference between asymmetry measures is not an issue for any interpretations advanced in this work; the result simply means that the measure of asymmetry from \textbf{\textit{FLEX}} is conceptually different from rotational asymmetry. Work following similar exercises to \citet{Conselice_(2003)}, e.g. using local galaxies as a calibration, may be a useful future study.

\section{Mock image measurements}
\label{app: F}

\begin{figure}
	\includegraphics[width=\columnwidth]{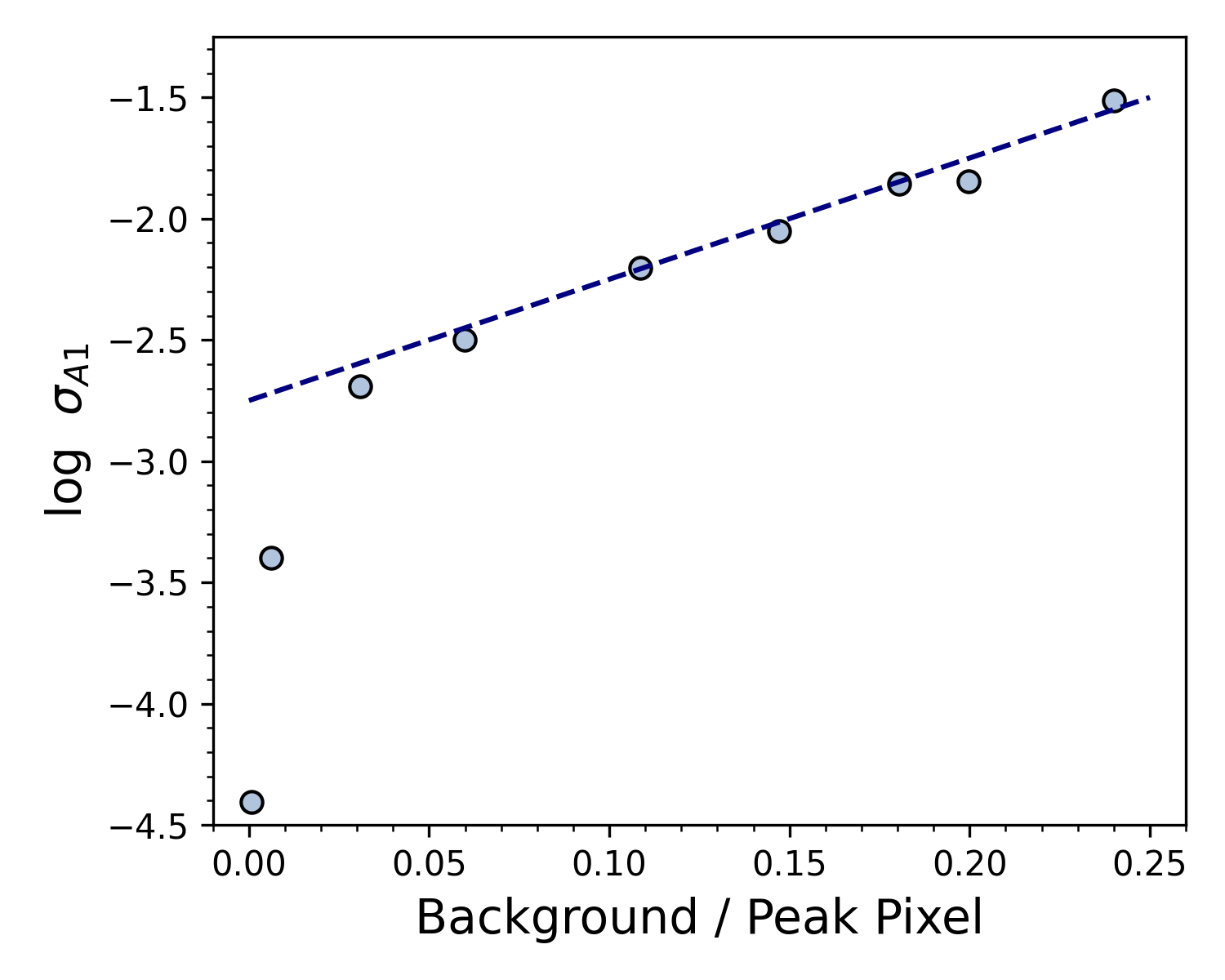}
    \caption{Empirical calibration of the effect of increasing background noise (measured as the background level divided by the peak pixel in the image) on the uncertainty estimating $A_1$ ($\log_{10} \sigma_{A_1}$). The uncertainty floor from centring (cf. Appendix~\ref{app: C}) is $\log_{10} \sigma_{A_1}=-2$, such that the background is only a significant source of uncertainty when S/N$~\lesssim6$. This is only true for a small handful of galaxies in our primary sample, in specific filters.}
    \label{fig:peak_pixel_significance}
\end{figure}

\begin{figure}
	\includegraphics[width=\columnwidth]{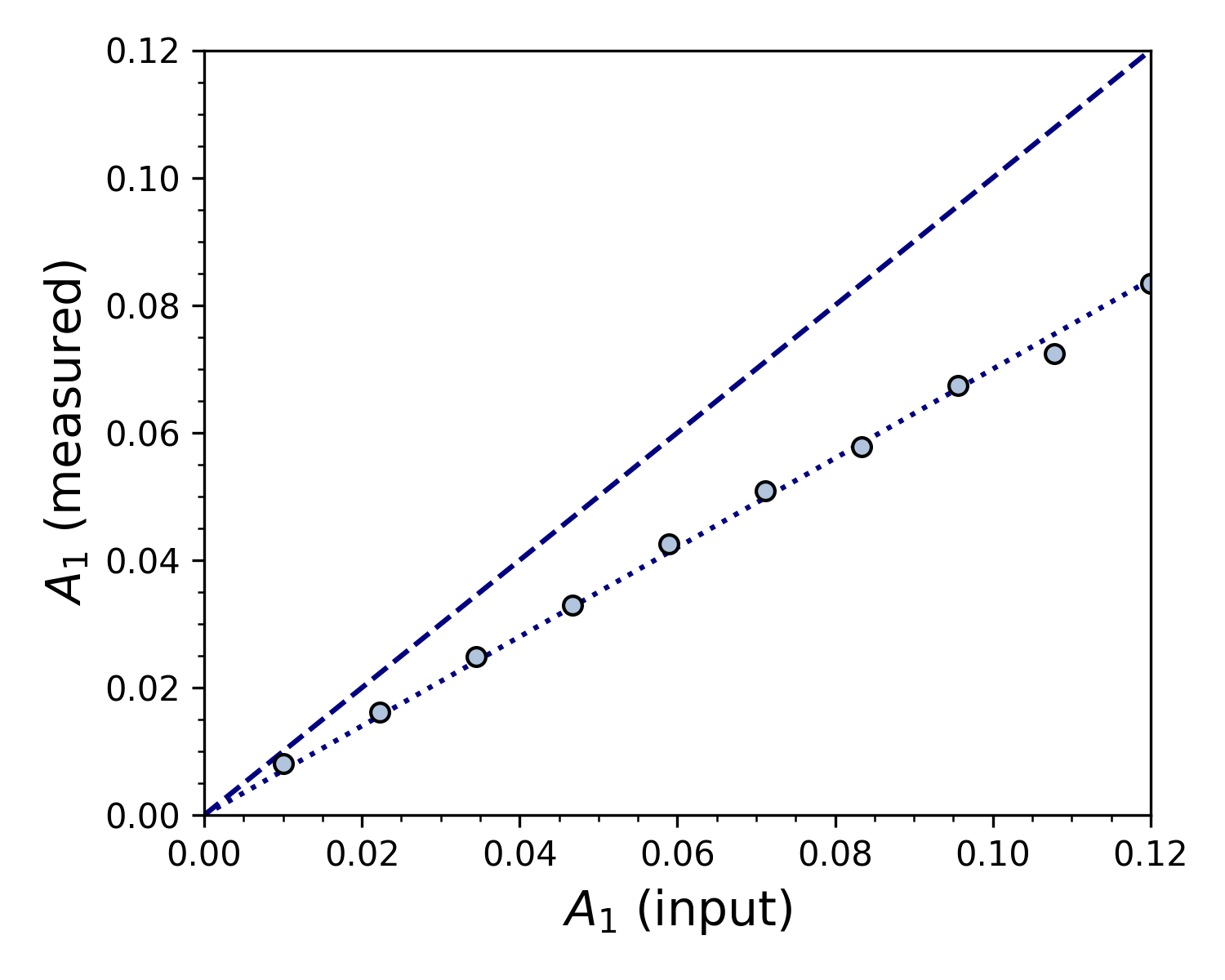}
    \caption{Recovery of asymmetry input to a model galaxy. We realise asymmetric galaxies as in the text, and measure the recovery by \textbf{\textit{FLEX}}. The dashed line indicates perfect recovery; we find that we recover 70 per cent of the input asymmetry in this test (dotted line).}
    \label{fig:injection_test}
\end{figure}

To demonstrate the utility and fidelity of \textbf{\textit{FLEX}} for measuring asymmetries, we perform two tests. First, measurement on an unperturbed galaxy to determine an uncertainty floor on $A_1$ from images with varying levels of observational noise, and second, an injection test, where we purposely deform a galaxy and recover the deformation.

In the first test, we realise an exponential disc of points (stars) and then `observe' the stars with varying pixel resolutions and increasing noise levels (as a function of peak surface brightness of the disc). We realise the exponential disc by drawing 10$^6$ `stars' from an exponentially distributed cumulative distribution function in radius and assigning a random azimuth to each star\footnote{We find empirically that 10$^6$ samples is sufficient to reduce $A_1$ measurement noise below the centring-induced uncertainty threshold of $\sigma_{A_1}=0.01$. Smaller sample numbers may create measurable $A_1$ signal.}. The galaxy is then `observed' by making a 2D histogram of the stars. The asymmetry measurement in this case should be (nearly) zero, as the galaxy is realised to be (near) perfectly symmetric. Any deviation from zero is the result of observing the mock galaxy, both in resolution and in image uncertainty. We find that in general, the measurement uncertainty of $A_1$ increases with increasing ratio of the background noise to the peak pixel, without increasing bias (Figure~\ref{fig:peak_pixel_significance}). High background (relative to the central galaxy) can affect $A_0$, which in turn will increase $A_1$ (which uses $A_0$ as a normalisation term). We use our model exponential disc to estimate the uncertainty contribution from $A_0$ by testing noise injection. Noise is injected into an image at varying levels with respect to the original galaxy image, and then we measure the results. The contribution of background measurement noise to $A_1$ uncertainty becomes significant (relative to other sources of uncertainty) at a ratio of background-to-peak pixel of 0.15, or (inverted to find) peak image S/N~$\approx 6$. We use an empirically calibrated uncertainty measure, $\log_{10}\left(\sigma_{A_1}^{\rm background}\right) = 5\left(\frac{{\rm background}}{{\rm peak~pixel}}\right) -2.75$. This is shown in Figure~\ref{fig:peak_pixel_significance}. The effect of the background on the calculation of $A_0$ is not reflected in the formal image uncertainties, as the image uncertainty does not create bias in $A_0$, and as such, the Monte Carlo realisations of $A_1$ (e.g. as described in Appendix~\ref{app: B}) do not vary strongly. This contribution to the overall uncertainty dominates the shorter wavelength points in Figure~\ref{fig:main_plot_1}, where the HST images have significantly less flux overall compared to their background (cf. Figure~\ref{fig:expansion_grid}). Note that at small levels of injected noise -- including when no noise in injected -- we find a very stable calculation of $A_1$, well below our approximated uncertainty measure.

We test the effect of pixel resolution by varying the pixels used to `observe' the model exponential disc, finding that once the uncertainty from the background level has been accounted for, the effect of pixel resolution is subdominant to other uncertainty terms, contributing at approximately the 0.05 per cent level.

In the second test, we again use the mock exponential galaxy. We first compute the expansion values and zero all but the contribution from the lowest-order term. We then set a value of one $m=1$ coefficient, and use this to create an asymmetric galaxy image by drawing 10$^6$ random `stars' from the specified surface density\footnote{Using {\tt lintsampler} \citep{Naik.lintsampler.2024}.} and observing them by pixelating the samples. 

We find that the recovery of the input $A_1$ value follows a monotonic trend, which confirms the ability of \textbf{\textit{FLEX}} to recover asymmetry trends. The recovery is at a consistently lower value than the input, which we attribute to imperfect sampling of the mock galaxy reconstruction. We empirically estimate to be 30 per cent from a linear fit to the data. We do not expect this significant of a bias in the \textbf{\textit{FLEX}}-derived measurements for real galaxies, and emphasise that the trend is the important quality to recover. In Figure~\ref{fig:injection_test}, we show the empirical recovery of $A_1$ as a function of input $A_1$, demonstrating our findings.


\bsp	
\label{lastpage}
\end{document}